\documentclass[sn-mathphys,Numbered]{sn-jnl}

\usepackage{graphicx}%
\usepackage{multirow}%
\usepackage{amsmath,amssymb,amsfonts}%
\usepackage{amsthm}%
\usepackage{mathrsfs}%
\usepackage[title]{appendix}%
\usepackage{xcolor, xurl}%
\usepackage{textcomp}%
\usepackage{manyfoot}%
\usepackage{booktabs}%
\usepackage{algorithm}%
\usepackage{algorithmicx}%
\usepackage{algpseudocode}%
\usepackage{listings}%
\usepackage{natbib} 
\usepackage[utf8]{inputenc}
\usepackage{hyperref}
\usepackage{etoolbox}
\usepackage{nicematrix}
\usepackage{tikz}
\usepackage{subfig}
\usepackage{unicode, tabularx}
\usepackage[title]{appendix}%
\usepackage{bm}
\usepackage{tabulary}

\newcommand{\etal}{\emph{et al.}}
\newcommand{\STAB}[1]{\begin{tabular}{@{}c@{}}#1\end{tabular}}

\def\bX{{\mathbf X}}

\def\bx{{\mathbf x}}
\def\by{{\mathbf y}}

\begin{document}

\title{Equipping Computational Pathology Systems with Artifact Processing Pipelines: A Showcase for Computation and Performance Trade-offs}

\author[1]{Neel \sur{Kanwal}*}
\equalcont{These authors contributed equally to this work.}

\author[2]{Farbod \sur{Khoraminia}}
\equalcont{These authors contributed equally to this work.}

\author[3,4]{Umay \sur{Kiraz}}
\equalcont{These authors contributed equally to this work.}

\author[5]{Andr\'{e}s \sur{Mosquera-Zamudio}}
\equalcont{These authors contributed equally to this work.}

\author[5]{Carlos \sur{Monteagudo}}

\author[3,4]{Emiel A.M. \sur{Janssen}}

\author[2]{Tahlita C.M. \sur{Zuiverloon}}

\author[1]{Chunming \sur{Rong}}

\author[1]{Kjersti \sur{Engan}}

\affil[1]{Department of Electrical Engineering and Computer Science, University of Stavanger, 4021 Stavanger, Norway}
\affil[2]{Department of Urology, University Medical Center Rotterdam, Erasmus MC Cancer Institute, 1035 GD Rotterdam, The Netherlands}
\affil[3]{Department of Pathology, Stavanger University Hospital, 4011 Stavanger, Norway}
\affil[4]{Department of Chemistry, Bioscience and Environmental Engineering, University of Stavanger, 4021 Stavanger, Norway}
\affil[5]{Department of Pathology, INCLIVA Biomedical Research Institute, and University of Valencia, 46010 Valencia, Spain\\
* Corresponding author(s): \{neel.kanwal, kjersti.engan\}@uis.no}

\abstract{
\textbf{Background: } Histopathology is a gold standard for cancer diagnosis. It involves extracting tissue specimens from suspicious areas to prepare a glass slide for a microscopic examination. However, histological tissue processing procedures result in the introduction of artifacts, which are ultimately transferred to the digitized version of glass slides, known as whole slide images (WSIs). Artifacts are diagnostically irrelevant areas and may result in wrong predictions from deep learning (DL) algorithms. Therefore, detecting and excluding artifacts in the computational pathology (CPATH) system is essential for reliable automated diagnosis.

\textbf{Methods:} In this paper, we propose a mixture of experts (MoE) scheme for detecting five notable artifacts, including damaged tissue, blur, folded tissue, air bubbles, and histologically irrelevant blood from WSIs. First, we train independent binary DL models as experts to capture particular artifact morphology. Then, we ensemble their predictions using a fusion mechanism. We apply probabilistic thresholding over the final probability distribution to improve the sensitivity of the MoE. We developed four DL pipelines to evaluate computational and performance trade-offs. These include two MoEs and two multiclass models of state-of-the-art deep convolutional neural networks (DCNNs) and vision transformers (ViTs). These DL pipelines are quantitatively and qualitatively evaluated on external and out-of-distribution (OoD) data to assess generalizability and robustness for artifact detection application.

\textbf{Results: } We extensively evaluated the proposed MoE and multiclass models. DCNNs-based MoE and ViTs-based MoE schemes outperformed simpler multiclass models and were tested on datasets from different hospitals and cancer types, where MoE using (MobileNet) DCNNs yielded the best results. The proposed MoE yields 86.15 \% F1 and 97.93\% sensitivity scores on unseen data, retaining less computational cost for inference than MoE using ViTs. This best performance of MoEs comes with relatively higher computational trade-offs than multiclass models. Furthermore, we apply post-processing to create an artifact segmentation mask, a potential artifact-free RoI map, a quality report, and an artifact-refined WSI for further computational analysis. During the qualitative evaluation, field experts assessed the predictive performance of MoEs over OoD WSIs. They rated artifact detection and artifact-free area preservation, where the highest agreement translated to a Cohen kappa of 0.82, indicating substantial agreement for the overall diagnostic usability of the DCNN-based MoE scheme.

\textbf{Conclusions: }
The proposed artifact detection pipeline will not only ensure reliable CPATH predictions but may also provide quality control. In this work, the best-performing pipeline for artifact detection is MoE with DCNNs. Our detailed experiments show that there is always a trade-off between performance and computational complexity, and no straightforward DL solution equally suits all types of data and applications. The code and HistoArtifacts dataset can be found online at \href{https://github.com/NeelKanwal/Equipping-Computational-Pathology-Systems-with-Artifact-Processing-Pipeline}{Github} and \href{https://zenodo.org/records/10809442}{Zenodo}, respectively.
}


\maketitle
\section{Introduction} \label{sec:intro}

Cancer develops in organs when genetic mutations in normal cells trigger their transformation into tumor cells. This transformation may be triggered by frequent exposure to carcinogens, a class of substances (chemical, biological, or physical), or several other factors that have the potential to cause cancer~\cite{cancer-gov}. Diagnosing cancer accurately and efficiently is critical for medical treatment and a reduced mortality rate, given its status as one of the deadliest diseases worldwide, with a projected estimate of 29 million deaths by 2040~\cite{wcrf_stats, pulumati2023technological}. Histopathology is considered a gold standard for identifying cancerous cells, which involves examining tissue samples under a microscope using a histological glass slide~\cite{khened2021generalized}. However, this manual inspection and laboratory procedure is not without its pitfalls, as it is labor-intensive, subjective, and can be affected by inter- and intra-observer variability~\cite{zhu2019breast,  kanwal2023detection}.
Furthermore, the projected rise in cancer cases and the shortage of pathologists are significant issues that may lead to delayed diagnosis and treatment, resulting in a severe impact on clinical decision-making~\cite{car2016clinician}. Therefore, streamlining the traditional diagnostic process through digitization and automation can provide timely diagnosis, improved treatment decisions, and efficacy~\cite{pulumati2023technological}. Digital pathology (DP) has the potential to overcome these challenges by providing rapid diagnosis and smooth sharing of secondary opinions~\cite{pallua2020future}. In fact, in the last decade, there has been a five-fold growth in DP research and development~\cite{dimensionsai, kanwal2022devil}. This increase in the adoption of DP in clinical practice enables computation over the digitized version of histological slides, commonly called whole slide images (WSIs).

\begin{figure}[ht!] 
\centering
\includegraphics[width=1.01\linewidth]{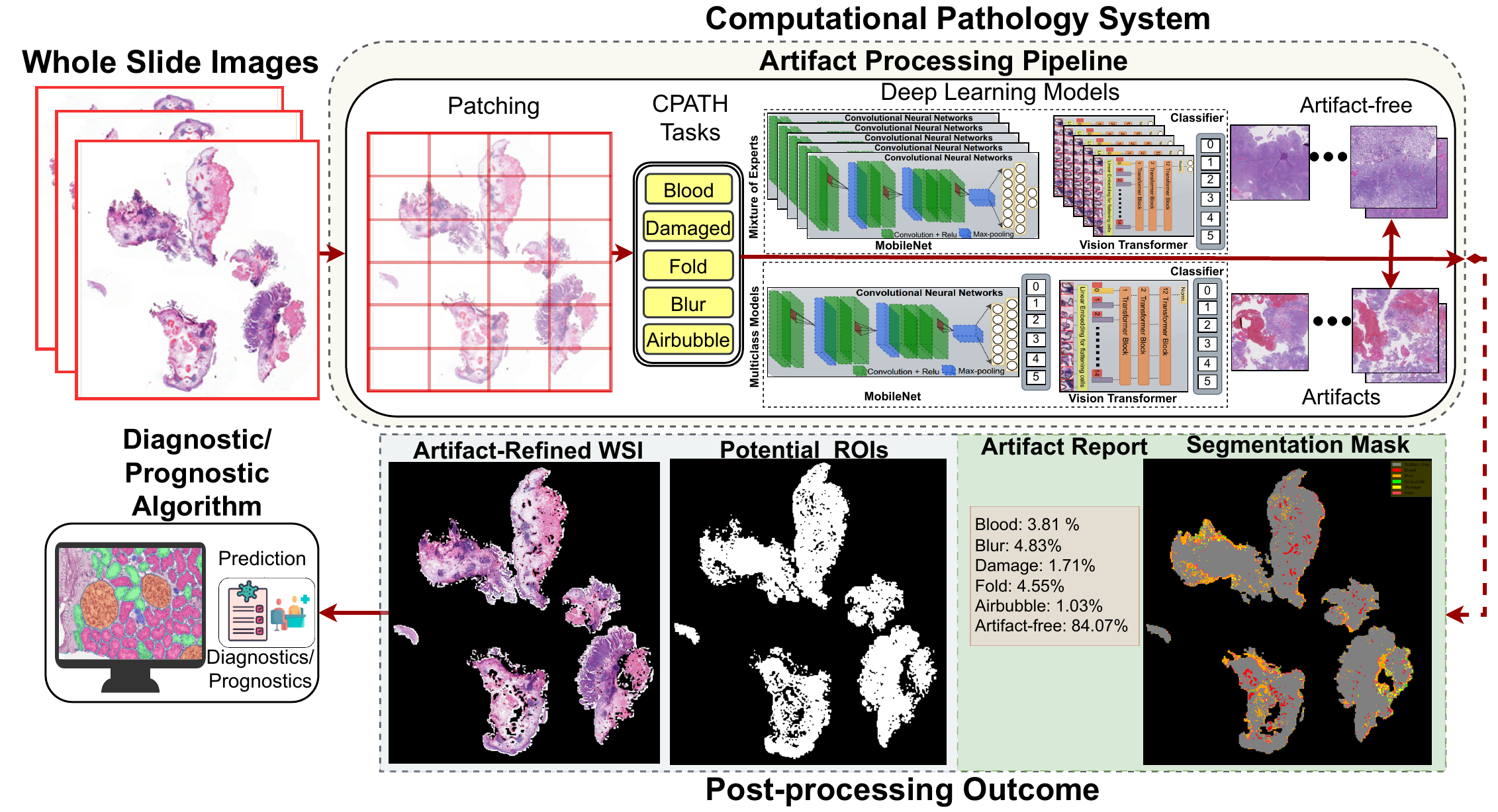}
\caption{\textbf{An overview of computational pathology (CPATH) system equipped with artifact processing pipeline.} Whole slide images (WSIs) are split into small sub-images (patches) to make them computationally tractable for deep learning (DL) models. These patches are fed to a mixture of experts (MoE) or multiclass models composed of state-of-the-art DL architectures to perform different CPATH classification tasks. Only patches with histological relevance can flow further for the downstream tasks. Finally, predictions are post-processed to produce different outcomes, such as a segmentation map, artifact report for quality control, region of interest mask, and artifact-free WSI for the diagnostic or prognostic algorithm to make a final clinical prediction.}
\label{main}
\end{figure} 

Computational pathology (CPATH) systems have the potential to unfold information embedded in WSIs by automated systems based on AI and image processing ~\cite{CAMPANELLA2018142,kanwal2022devil,hosseini2024computational}. The seamless integration of CPATH with DP can enhance diagnostic or prognostic methodologies and save pathologists' time~\cite{kanwal2023detection, louis2014computational}. However, artifacts that appear during the histological slide preparation are ultimately transferred to the WSIs~\cite{taqi2018review,bindhu2013facts,kanwal2023vision}. Artifacts are diagnostically irrelevant areas, and pathologists usually ignore these areas during manual inspection, but unfortunately, the presence of histological artifacts can hamper the performance of CPATH systems during automated diagnosis~\cite{kanwal2022devil, wright2020effect}. Therefore, it is essential to equip the CPATH system with an \emph{artifact detection pipeline} to exclude artifacts and ensure the flow of histologically relevant tissue for diagnostic or prognostic algorithms, as illustrated in Figure~\ref{main}. Thus, a CPATH system with artifact processing capacity will not only increase the likelihood of reliable and accurate predictions but also provide quality control (QC) for laboratory procedures, identifying weaknesses during the histotechnical stages (see review~\cite{kanwal2022devil}) in acquiring WSIs.

In recent years, deep learning (DL) approaches have garnered more attention from the CPATH community due to their ability to extract hidden patterns in histological data~\cite{tabatabaei2022residual,chen2022classification,9816352, litjens2017survey}. Popular DL architectures such as deep convolution neural networks (DCNNs) and vision transformers (ViTs) have widely been used as state-of-the-art (SOTA) to distinguish tissue patterns for different cancer types and perform image classification and segmentation tasks~\cite{kanwal2023vision,chen2022classification,chen2021review}. Some researches~\cite{lu2022bridging,zhu2023understanding} demonstrate that DCNNs perform better on small datasets, thanks to the inductive bias, which helps them to learn spatial relevance effectively. While other works~\cite{atabansi2023survey,naseer2021intriguing,bhojanapalli2021} argue in favor of ViTs, showing that they are highly robust, attend to overall structural information, and are less biased towards textures. Nevertheless, both DL architectures may suffer from overfitting, poor generalization, and reproducibility issues, leading to overconfident predictions on new (external) data. To address these problems, ensembles of DL models (a.k.a. deep ensembles) have been used to overcome the weakness of an individual model~\cite{hsu2022automatic,meng:2021,abe2022deep}. Ensemble methods combine the prediction of independent models using averaging or majority voting. A mixture of experts (MoE) is an extended method that trains DL for a sub-task and then combines the predictions dynamically to obtain a nuanced prediction. In short, the MoE approach consists of multiple DCNNs or ViTs, experts on each subclass, to achieve improved results. MoEs benefit in terms of reproducibility by reducing the variance of predictions but augmenting computational expense~\cite{mohammed2023comprehensive}. In contrast, the multiclass approach can be computationally efficient but does not involve the strength of multiple models, which are adaptive for looking into different aspects of data. Based on these arguments, the choice between DL approaches depends on application requirements. This raises a fundamental question: \emph{how to build an effective artifact detection DL approach for CPATH systems with suitable trade-offs between computational complexity and performance?}

An effective DL approach for artifact detection applications (our case) might be created using MoEs, one DL model for each artifact class, or multiclass models with multiple output classes. In this paper, we propose the MoE-based DL approach, which uses a fusion mechanism to integrate predictions from experts and apply probabilistic thresholding to improve the sensitivity. We establish several DL pipelines using the MoE and multiclass models for detecting notable artifacts (i.e., damaged tissue, blur, folded tissue, air bubbles, and diagnostically irrelevant blood) from histological WSIs (see Figure~\ref{main}). Our DL pipelines produce four outcomes for the input WSI: i) Artifact segmentation map; ii) Artifact report for QC using six classes (five artifacts and artifact-free area); iii) Artifact-free mask with potential regions of interest (RoIs) with diagnostic relevance; and iv) Artifact-refined WSI for the diagnostic algorithm. 

Our contributions to this work are summarized below:

\begin{itemize}[wide]

    \item We develop four DL models (referred to as DL pipeline throughout the paper), with SOTA DCNNs (MobileNet~\cite{mobilenet}) and ViTs (ViT-tiny~\cite{Deit}), using MoE and a multiclass approach. 
    
    \item We evaluate the computational complexity of the pipelines and systematically choose a learned probability threshold for maximizing the sensitivity of DL models in external validation.
    
    \item We conduct a qualitative and quantitative evaluation over external data (from different cancer types) and assess the efficiency of the proposed MoE scheme for detecting artifacts and QC.
\end{itemize}

The paper is structured as follows: Section~\ref{sec:related} presents recent studies involving DL approaches for computational pathology and related work for detecting artifacts. Section~\ref{sec:data_material} provides data material descriptions.
Section~\ref{sec:methods} explains pre-processing for creating datasets, the proposed method, post-processing, evaluation metrics, and implementation details. Section~\ref{sec:results} discusses results for performance and computational complexity. Section~\ref{sec:conclusion} concludes this work. Finally, Section~\ref{sec:limitations} discusses limitations and future directions for a smooth integration of artifact processing pipelines in CPATH systems.

\section{Related Work} \label{sec:related}

\subsection{Deep Learning for Computational Pathology}
Deep learning (DL) approaches have gained popularity in the CPATH community for different tasks~\cite{morales2021artificial, bulten2022artificial, khoraminia2023artificial, litjens2017survey}. In recent years, several works~\cite{gay2019texture, GANDOMKAR201814, 9816352, wessels2023self, stegmuller2023scorenet} have used popular DL architectures for diagnosis and prognostic algorithms. FDA-approved PAIGE~\cite{perincheri2021independent} is an example of such a DL-based algorithm for prostate cancer. These works can be roughly divided into two branches, such as DCNN-based (MobileNet~\cite{mobilenet}, DenseNet~\cite{huang2017densely}, ResNet~\cite{resnet}, or GoogleNet~\cite{googlenet}, etc.) or ViT-based (ViT-Tiny(~\cite{Deit}, DINO~\cite{caron2021emerging}, or SwinTransformer~\cite{zidan2023swincup} etc.) approaches. 

In the first branch, Srinidhi \etal~\cite{srinidhi2020deep} comprehensively reviewed different DL approaches for developing disease-specific classification algorithms using histological images. Riasatian \etal~\cite {RIASATIAN2021102032} applied transfer learning over DCNNs to classify various tumor types and accomplished remarkable results using three public histopathology datasets. Talo~\citep{TALO2019101743} demonstrated that pre-trained ResNet~\cite{resnet} and DenseNet~\cite{huang2017densely} achieved better accuracy than traditional methods in the literature for classifying grayscale and color histopathological images. Similarly, Wang \etal~\citep{wang:2020} proposed a DCNN-based method based on GoogleNet~\cite{googlenet} to locate tumors in breast and colon images using complex example-guided training for WSI analysis. Among other DCNN works, Meng \etal~\citep{meng:2021} compared several architectures for classification and segmentation problems on a cervical histopathology dataset. Their approach found the best results for precancerous lesions using ResNet-101~\cite{resnet}. For the same task, MobileNet~\cite{mobilenet} was the fastest. Wang \etal~\cite{WANG2022103451} performed multi-class breast cancer classification in their two-stage dependency-based framework. A MobileNet~\cite{mobilenet} was used as a backbone to extract the features in the first stage. Then, the MobileNet~\cite{mobilenet} backbone was modified to perform sub-type classification for benign and malignant categories. Gandomkar \etal~\cite{GANDOMKAR201814} deployed ResNet~\cite{resnet} for classifying breast histology images into benign or malignant and then identified them among several sub-types using a meta-classifier based on a decision tree. 

Works in the second branch used ViTs, which have emerged as new SOTA, leveraging attention mechanisms to improve shape understanding and generalizablity~\cite{naseer2021intriguing}~\cite{bhojanapalli2021}. Stegmüller \etal~\cite{stegmuller2023scorenet} developed ViT-based ScoreNet for breast cancer classification. Their approach attended to some regions in the WSI for faster processing based on image semantics. Wessel\etal~\cite{wessels2023self} used DINO~\cite{caron2021emerging} for predicting overall and disease-specific survival in renal cell carcinoma. Zidan \etal~\cite {zidan2023swincup} introduced a ViT-based cascaded architecture for segmenting glands, nuclei, and stroma in colorectal cancer. Gao \etal~\cite{gao2021instance} proposed instance-based ViT to capture global and local features for subtyping papillary renal carcinoma, achieving better performance over selected RoIs. 

Unsurprisingly, in both branches, most of these DL algorithms were trained and tested on manually annotated clean data (with diagnostic relevance) and overlooked the impact of potential noise (histological artifacts) during the inference stage or unseen scenarios. 

Schomig \etal~\cite{schomig2021quality}, in their stress-testing study, showed that the accuracy of the prostate cancer DL algorithm was negatively affected by the presence of artifacts and resulted in more false positives. Even the presence of artifacts in the training data may result in poor learning by DL models, as they add irrelevant features to the data~\cite{kanwal2022devil,linmans2024diffusion}. Wright \etal~\cite{wright2020effect} demonstrated that removing images with artifacts improved the accuracy of DL models by a significant margin. Laleh \etal~\cite{ghaffari2022adversarial} emphasized the need for robustness of DL-based CPATH systems against artifacts for their widespread clinical adaptability. Artifact processing pipeline that can detect, extract, and eliminate non-relevant patches from WSIs before running a diagnostic algorithm would avoid any detrimental effect on downstream image analysis~\cite{CAMPANELLA2018142, wright2020effect, kanwal2024extract}. Therefore, it is essential to equip CPATH systems with artifact detection ability, which is also the focus of this work, to obtain reliable predictions~\cite{kothari2013eliminating, wright2020effect, kanwal2024sure}. 

\subsection{Detection of Histological Artifacts}

Most researches focus on reducing color variations and image augmentations during the preprocessing phase in CPATH literature~\cite{salvi2021impact, Perez-Bueno2020a}. Detection of artifacts is often an underrepresented aspect of WSI pre-processing~\cite{kanwal2022devil}. Compared to color normalization techniques, there remains a scarcity of research detecting notable artifacts before feeding histologically relevant data to the diagnostic algorithms. While some works~\cite{wright2020effect, Ameisen2014, shrestha2016, bahlmann2012automated} have relied on quickly identifying faulty WSIs by doing QC at low magnification. Avanki \etal~\cite{Avanaki2016} proposed a quality estimation method by combining blurriness, contrast, brightness, etc., to accept or discard WSI based on a reference. HistoQC~\cite{histoqc} provides content-based evaluation for finding outliers in a cohort of WSIS. Bahlmann \etal~\cite{bahlmann2012automated} exploited texture features and stain absorption to separate diagnostically relevant and irrelevant regions. However, artifacts appearing in diagnostically relevant areas are likely to be missed. Apart from their limitations with lower magnification, they were validated based on specific staining and tissue types. Therefore, artifact detection methods need to be extended to higher magnification. Moreover, artifact detection methods that can identify specific artifacts are desirable for QC, as some artifacts, like a blur, can be avoided by re-scanning glass slides or de-blurring techniques.

Earlier works for artifact detection relied on traditional image processing and color-space transformation approaches. Gao \etal~\cite{gao2010automated} detected blurry areas using 44 handcrafted (local statistics, brightness, etc) features. Hashimoto \etal~\cite{hashimoto2012} combined image sharpness and noise information to create a regression model for out-of-focus detection. For folded tissue detection, Palokangas \etal~\cite{palokangas2007segmentation} transformed red, green, and blue (RGB) images to hue, saturation, and intensity (HSI) to apply k-means clustering over the different saturation and intensity values. Bautista and Yagi~\cite{Bautista2009} detected folds using RGB shift with fixed thresholding on luminance and saturation values to enhance color structure in thick (folded) areas. Kothari \etal~\cite{kothari2013eliminating} introduced a rank-sum approach that used connectivity descriptors and image features to discard folded tissues. Their approach used two adaptive thresholds on saturation and intensity ranges. Chadaj \etal~\cite{swiderska2016automatic} separated uninformative blood (hemorrhage) from blood vessels using cyan, magenta, yellow, and black (CYMK) color space and morphology. Mercan \etal~\cite{mercan2014localization} proposed a k-means method to classify diagnostically relevant vs. non-relevant patches using local binary patterns extracted from stains and L*a*b histograms. A detailed review of other artifact detection works can be found in Kanwal \etal~\cite{kanwal2022devil}. Since color-based approaches can heavily underperform when exposed to data from different cohorts with stain variation, data-driven DL approaches are needed to resolve the challenges.

Among recent works using DL-based approaches, Albuquerque \etal~\cite{albuquerque2021deep} compared several DCNNs for detecting out-of-focus areas in their ordinal classification problem. Kohlberger \etal~\cite{kohlberger2019whole} proposed ConvFocus to quantify and localize blurry areas in WSI. Wetteland \etal~\cite{Wetteland2019, wetteland2020multiscale} proposed a segmentation model to find blood and damaged tissue in bladder cancer WSIs. Clymer \etal~\cite{clymer2020decidual} developed a two-stage method to detect blood at low resolution using RetinaNet and later Xception CNN for subsequent classification. Babie \etal~\cite{babaie2019deep} used SOTA DCNNs with SVM, KNN, and decision tree classifiers to separate folded tissues from normal tissue in a binary fashion. Kanwal \etal~\cite{kanwal2022quantifying} used several DCNNs to assess the impact of color normalization over blood and damaged tissue detection. In another work~\cite{kanwal2023vision}, they trained ViT-Tiny~\cite{Deit} for air bubble detection using knowledge distillation. All these works relied on training a single network to classify one or two artifacts against an artifact-free class. It is a well-known problem that DL models suffer from poor generalization, robustness, and overconfident predictions over out-of-distribution (OoD) data~\cite{guo2017calibration,linmans2023predictive, zhang2021understanding}. Thus, the high variance in the prediction of DL models needs to be addressed, especially when deployed in a critical application. A prominent DL technique, "deep ensembles," resolves these problems by training several baseline DL architectures and combining the resultant predictions to increase accuracy and OoD performance~\cite{zhu2019breast}. However, the success of the ensemble method relies on several factors, such as how baseline models are trained and integrated. The most widely used ensemble techniques include averaging and majority voting~\cite{mohammed2023comprehensive}. It is worth noting that a simple aggregation using averaging methods or majority voting is not a smart choice and is very sensitive to biased baseline models~\cite{mohammed2023comprehensive}.  

A mixture of experts (MoE) may address this shortcoming by combining base learners, who are experts on detecting particular artifact morphology. Unlike deep ensemble, where all models are trained on the same data, in MoE, each DL model is trained for a sub-task to master specific aspects of the data, resulting in improved robustness. This is the first work to provide a comprehensive DL-based artifact processing pipeline that takes the entire WSI, preprocess, infer, and post-process in an end-to-end fashion for artifact detection and QC applications.

\section{Data Materials}
\label{sec:data_material}
This section details the histological data used for training and validating DL models. The following in-house (private) datasets are used for the experiments.

\subsection{Training and Development Data}\label{dataset:emcdev}
We obtained 55 WSIs of bladder cancer resection biopsies from the Erasmus Medical Centre (EMC) in Rotterdam, The Netherlands. The glass slides were formalin-fixed, stained with Hematoxylin (purple) and Eosin (pink) (H\&E) dyes, scanned with a Hamamatsu Nanozoomer 2.0HT at 40× and saved in \emph{ndpi} format with a pixel size of 0.227~{$\mu$}m $\times$ 0.227~{$\mu$}m. These WSIs were properly anonymized to preserve patient privacy, and all ethical requirements were followed before the dataset was created. \textcolor{blue}{NK received training for the task and manually annotated five artifacts (blurry areas, folded tissues, blood (hemorrhage), air bubbles, and damaged tissue). The rest of the tissue was marked as an artifact-free region. Note that not all WSIs contained five artifacts present and they were not extensively labeled as distinct tissue types since this histological data is not used for any task other than artifact detection. However, each WSI had at least one annotation for RoI or the artifact region (i.e., blur, fold, etc).} Later sections refer to this cohort as \emph{EMC}$_{dev}$. A detailed description of the prepared dataset and its availability is mentioned in Section~\ref{sec:preprocessing}.

\subsection{External Validation Data} \label{sec:external}
Since these external cohorts are not involved in training and development, they can be considered OoD data with different tissue types and staining. We have used the following cohorts for inference only to validate the generalizability and robustness of the proposed methods. 

\subsubsection{EMC Cohort:} \label{dataset:emc} This cohort is a collection of bladder cancer WSIs from a multi-center cohort provided by Erasmus MC, Rotterdam, The Netherlands. These WSIs with \emph{MRXS} format were prepared with H\&E staining and scanned with a 3DHistech P100 scanner at 80$\times$ magnification. A few WSIs were selected based on the presence of artifacts to test their generalization ability. \textcolor{blue}{FK manually annotated these WSIs for five artifacts, in a similar fashion as mentioned earlier.} We have used a 40$\times$ magnification level for inference as the models are trained at a similar level.  We will refer to this dataset as \emph{EMC}$_{inf}$, and it is a different cohort than the above-mentioned \emph{EMC}$_{dev}$.

\subsubsection{SUH Cohort:} \label{dataset:suh} This cohort is a private breast cancer cohort of 258 surgical specimens. It contains H\&E WSIs prepared from surgical specimens and collected by the Stavanger University Hospital (SUH) in Norway between 1978 and 2004. The WSIs are in \emph{ndpi} format and scanned using the Hamamatsu NanoZoomer S60 at 40$\times$ magnification. An expert pathologist (UK) selected and manually annotated a few WSIs based on the severity of the presence of these artifacts. Only five artifacts were carefully annotated, and the rest were marked as artifact-free regions. We have used these WSIs to test DL pipelines over cancer types that differ from the ones they are trained on. We will refer to this dataset as \emph{SUH}$_{inf}$.

\subsubsection{INCLIVA Cohort:} \label{dataset:inc}This cohort was prepared by the Department of Anatomical Pathology of the Hospital Clínico Universitario de Valencia, Spain. It is a collection between 1988 and 2020. The glass slides were prepared from skin cancer biopsies and were scanned with Roche's Ventana iScan HT at 40$\times$ magnification. WSIs were saved in \emph{tiff} format. An expert dermatopathologist (AM) selected and annotated a few WSIs with artifacts from this cohort to validate the proposed pipeline over the external cohort. We will refer to this dataset as \emph{INCLIVA}$_{inf}$.

\begin{table}[h]
  \caption{Breakdown of the number of patches obtained in each class of the dataset $\mathcal{D}$, obtained form \emph{EMC}$_{dev}$ after preprocessing.}
    \label{tab:dataset_overview}
    \centering
    \begin{tabular}{|l|c|c|c|c|}
         \hline
         \textbf{(label) Class} & \textbf{\shortstack{(35 WSIs)\\ Training}}  & \textbf{\shortstack{(10 WSIs)\\ Validation}} & \textbf{\shortstack{(10 WSIs)\\ Test}} & \textbf{Total} \\
         \hline  \hline
         \textbf{(0) Artifact-free } & 5,249 & 1,591 & 965  & 7,805 \\ 
         \textbf{(1) Blood } & 16,743 & 4,186 & 5996 & 26,655 \\ 
         \textbf{(2) Blur } & 5,661  & 754 & 1,137  & 7,552 \\ 
         \textbf{(3) Air bubbles } & 2,499  & 1,175 & 846   & 4520 \\ 
         \textbf{(4) Damaged Tissue } & 2,577  & 332 & 1,013  & 3,922 \\ 
         \textbf{(5) Folded Tissue } & 998  & 114 & 131  & 1,243 \\ 
         \hline
    \end{tabular}
\end{table}

\begin{figure}
    \centering
    \includegraphics[width=1.05\textwidth]{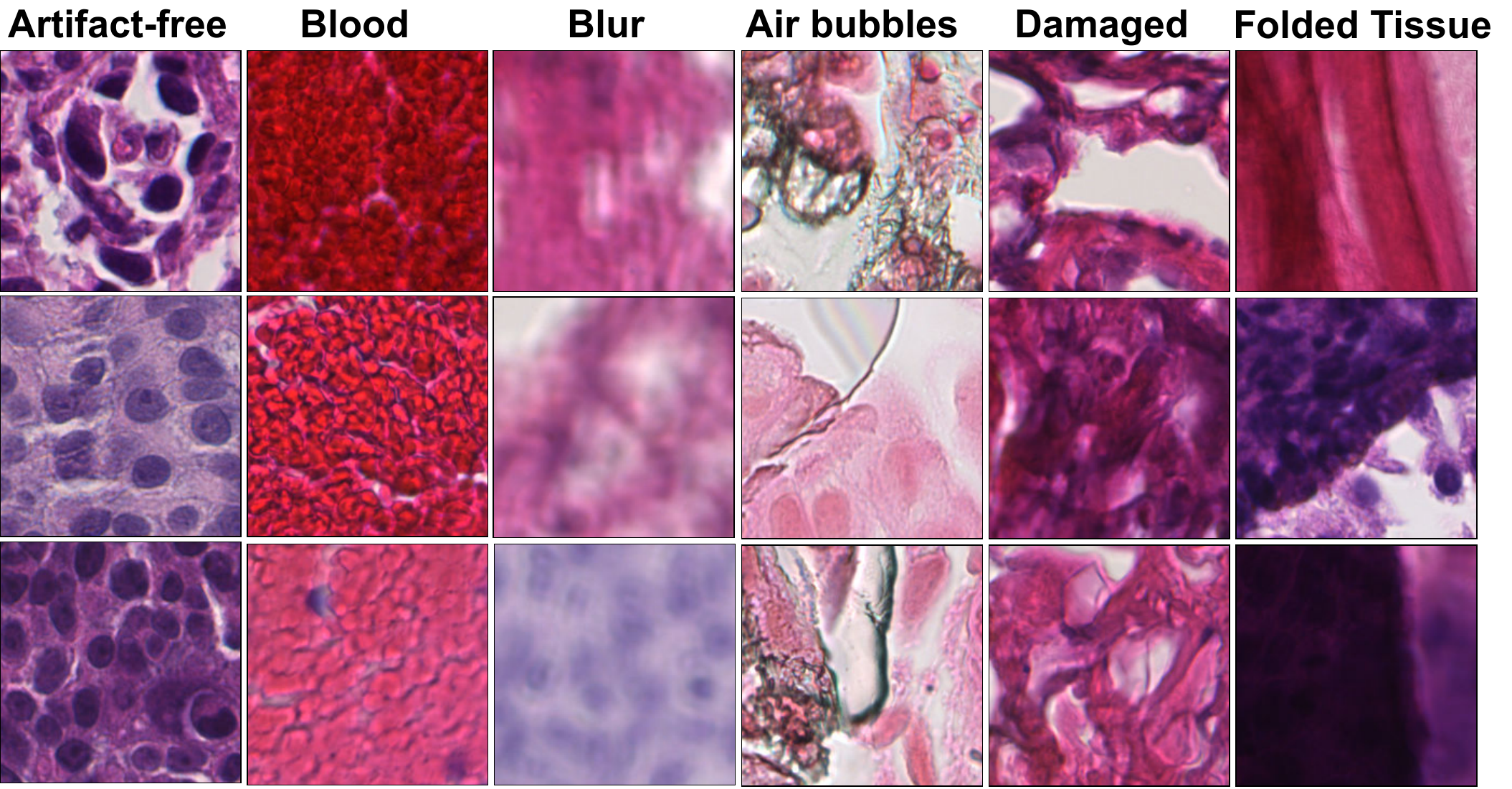}
    \caption{Examples of artifact-free and artifact-classes patches in our
prepared patch-based dataset $\mathcal{D}$ from \emph{EMC}$_{dev}$, and extracted at 40x magnification.}
    \label{fig:dataset}
\end{figure}

\section{Proposed Method} \label{sec:methods}
This section describes the data pre-processing, the proposed method for MoE, post-processing, evaluation metrics, and details of the implementation of the DL pipelines. 

Figure~\ref{main} gives a graphical overview of the proposed DL method for detecting histological artifacts in WSIs. We proceed with the artifact detection task in two steps. First, we train binary and multiclass models for patch-wise classification. The binary models are trained to detect one particular artifact, i.e., blur against artifact-free. The multiclass models provide output with six classes (five artifacts and one artifact-free). In the second step, we used these trained binary models to create a sort of MoE for inference and post-processing the predictions. We combine predictions from each expert in MoE by fusing their outputs. We apply a probability threshold to maximize sensitivity for detecting notable artifacts and providing artifact-free WSI with diagnostic potential. We deploy multiclass models with probabilistic thresholding similar to MoE. A detailed description of the proposed method is given below.

\subsection{Pre-processing} \label{sec:preprocessing}
We have used the $EMC_{dev}$ cohort to prepare the dataset. This included WSIs from this cohort, which were divided into 35/10/10 training, validation, and test WSIs to prevent data leakage. 

Let a WSI at magnification level 40$\times$ (sometimes known as 400$\times$) be denoted by $I_{\text{WSI}(i)}^{40\times}$ for specific $(i)^{th}$ WSI. Since $I_{\text{WSI}}^{40\times}$ are huge gigapixel images, it is not possible to process the entire WSI in compute memory at once. Most CPATH systems first tile or patch the WSI, or RoI, to make computation feasible before processing it further. The initial step in the patching procedure was to separate the foreground tissue from the background (white) areas irrelevant for image analysis. Foreground/background separation is usually done with a low-resolution version of the image, which can later be interpolated to be used with the full-resolution image. We obtained tissue foreground by transforming the RGB (red, green, and blue) color space to HSV (hue, saturation, and value). Later, Otsu thresholding was performed on the value channel to separate the foreground-containing tissue from the background. We set a uniform patch-coordinate sampling grid over the extracted foreground. Patches having at least 70\% overlap with the annotation area ($R$) were retrieved after the extracted foreground was tiled across the grid with a non-overlapping stride, as depicted in Figure~\ref{main}. 

Assuming ${\cal T} : I_{\text{WSI(i)}\in R}^{40\times} \rightarrow \{{\boldsymbol{x}}_{j}^{i}; j=1\cdots J \}$ denotes the patching process, which gives a set of $J$ patches over $R$. Here, $\bx_j^{i}\in\mathbb{R}^{W\times H \times C}$ corresponds to patch $j$ with coordinates $(x_{ij},y_{ij})$ from $\text{WSI}_{i}$ and H, W, and C represent the width, height, and channels of the patch, respectively. We refer to this prepared patch-based dataset from $EMC_{dev}$ as $\mathcal{D}=(\bX, \by) = \{(\bx_n,y_n)\}_{n=1}^{N}$ containing $N$ patches. Here $\bx_{n}$ is a vector and denotes $n$-th instance with $224\times224\times3$ pixels, and $y_n\in\{0,...,k\}$ is a scalar, where $k=1$ for binary and $k=5$ for multiclass dataset formulation. For instance, in a multiclass dataset, `0' represents artifact-free class, and \{1,2,3,4,5\} correspond to the blood, blur, air bubbles, damaged tissue, and folded tissue classes, respectively. Table~\ref{tab:dataset_overview} shows the breakdown of patches in each subset of the dataset $\mathcal{D}$ and Figure~\ref{fig:dataset} shows example instances for all classes obtained from $I_{\text{WSI}}^{40\times}$. This training and development dataset, named HistoArtifacts, is publicly available and can be downloaded from \href{https://zenodo.org/records/10809442}{Zenodo}.

\subsection{Feature Extractors and Classifiers} \label{sec:models}
The feature extractor and classifier are two significant components of most DL models for classification tasks. Feature extractors are crucial in DL algorithms as they help identify critical features in the data. In short, it reduces the dimensionality of the image and facilitates classification from a vector. Based on artifact detection works in the literature~\cite{kanwal2022quantifying,kanwal2023vision}, we have selected two popular DL architectures as feature extractors due to their smaller parametric size and faster inference: i) DCNN-based MobileNetv3~\cite{mobilenet} architecture, and ii) Vision transformer-based ViT-Tiny~\cite{Deit} architecture. 

\textbf{MobileNetv3:} MobileNetv3 is a SOTA DCNN architecture proposed by Howard\etal~\cite{mobilenet} and is part of the family of computationally efficient models for small devices by Google. The basic building blocks of MobileNetv3 include depth-wise separable convolutions and inverted residual blocks designed to reduce computational complexity and improve accuracy. MobileNetv3 is optimized through a combination of hardware-aware network architecture search and novel architecture advances, including the use of hard-swish activation and squeeze-and-excitation modules~\cite{mobilenet}. This architecture is released in different variants. The large architecture variant (used in this work) has a 5.4M parameter and is lightweight and efficient, making it suitable for computationally efficient image classification pipelines. 

\textbf{Vision Transformer:} Vision Transformers (ViTs) have gained attention as a new SOTA for image recognition tasks~\cite{bhojanapalli2021,naseer2021intriguing}. ViT architecture breaks down an input image into a series of smaller patches, linearly embeds each patch, adds position embeddings, and then feeds the resulting sequence of vectors to a standard Transformer encoder~\cite{dosovitskiy2020image}. This Transformer encoder consists of a stack of identical layers. It uses a self-attention mechanism to focus on different parts of the input by computing a weighted sum of the input features based on their similarity. We use a lightweight and efficient variant of the ViT architecture, ViT-Tiny~\cite{Deit}, with 6M parameters for faster inference.

We apply transfer learning to train DL models and update model parameters at each epoch. Assume $\phi$ represents our feature extractor with $\theta_{f}$ parameters. Then, for the input patch ($\bx_{n}$) with ground truth ($y_n$), we get a flattened feature embedding ($a_n$) using;
\begin{equation} \label{eq:emedding}
    \phi_{\theta_{f}} (\bx_n) =  a_{n}   \quad where \quad a_{n} = \{a_1,a_1,....,a_z\}
\end{equation}

For patch-wise classification, we train classifiers in a binary and multiclass fashion. We appended a three-layer fully connected (FC) classifier ($C_{\theta_{c}}$) at the end of the feature extractor. Let us denote our DL models with notation $\psi_{\theta}$, where $\theta=\theta_{f} \cup \theta_{c}$, denotes the parameter set of both the feature extractor and the classifier. To obtain the output probability vector ($\boldsymbol{P_{y_n}}$) for the input patch, we apply softmax ($\sigma$) to the output logits of the classifier as shown in Eq.~\eqref{eq:probab}. For instance, binary models predict (artifact vs. artifact-free), and multiclass models predict (five artifact classes vs. artifact-free), as shown in Eq.~\eqref{eq:classifier}.

\begin{equation}\label{eq:probab}
\centering
    \boldsymbol{P_{y_n}}(\bx_n) = \psi_{\theta}(\bx_n) = \sigma(C_{\theta_{c}}(\phi_{\theta_{f}}(\bx_n)))  = \sigma (C_{\theta_{c}}(a_n))  
\end{equation}

\begin{equation}\label{eq:classifier}
    \centering \large
    \boldsymbol{P_{y_n}}  = 
  \begin{cases}
    [p_{y_{0}}, p_{y_{1}}]^T &  \quad \textrm{if}  \quad \textrm{binary} \\
    [p_{y_{0}}, p_{y_{1}}, p_{y_{2}} , p_{y_{3}} , p_{y_{4}} , p_{y_{5}}]^T  &  \quad  \textrm{if}  \quad \textrm{multiclass}
   \end{cases}
\end{equation}

Here, $y_{p_0}$ is the probability of being an artifact-free class. In the binary model, $y_{p_1}$ corresponds to artifact class and in the multiclass model, $[y_{p_{1}}, y_{p_{2}}, y_{p_{3}}, y_{p_{4}}, y_{p_{5}}]$ are predicted probabilities for blood, blur, air bubbles, damaged tissue, and folded tissue classes respectively. Finally, we calculate cross-entropy loss between the ground truth and the prediction, back-propagate this loss, and update model parameters, $\theta$, at each epoch based on the experimental setup explained in Sec.~\ref{sec:experimental_detail}.


\begin{equation} \label{eq:loss} 
    L_{CE}(y_n,P_{y_n}) =
    \begin{cases}
    - y_n \cdot log(p_{y_0}) + (1-y_n) \cdot log (1-p_{y_0}) \qquad  \textrm{for binary} \\ \\
    - \sum_{i=0}^{k} y_n \cdot log(p_{y_i}) \qquad \qquad \quad \qquad \qquad \textrm{for multiclass}
    \end{cases}
\end{equation}

To obtain final predictions ($\hat{P}_{y_n}$) for classes, we apply argmax to $\boldsymbol{P_{y_n}}$.

\begin{align}\large
    \boldsymbol{\hat{P}_{y_n}} = argmax(\boldsymbol{P_{y_n}})
\end{align}

At the inference stage, we establish four DL pipelines using combinations of trained models, i.e., multiclass models (with MobileNetv3 and ViT-Tiny) and MoEs (combining binary MobileNetv3 and ViT-Tiny), as explained further in the following sections.

\begin{figure}
    \centering
    \includegraphics[width=0.7\textwidth]{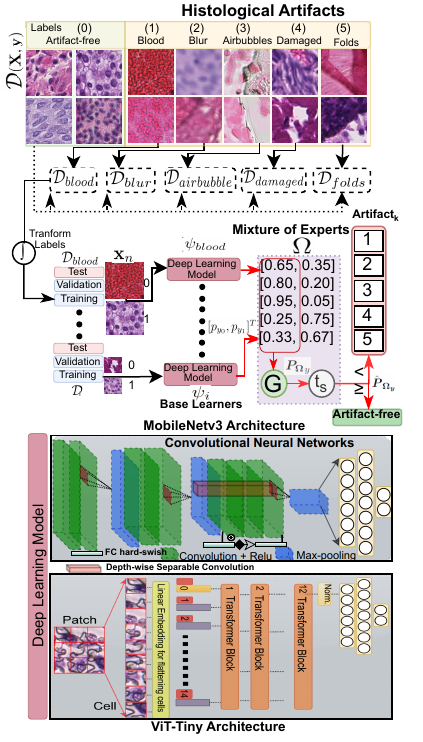}
    \caption{\textbf{An overview of the mixture of experts (MoE) formation for artifact detection}. Five base learners (either MobileNetv3 or ViT-Tiny deep learning architectures) are trained on overlapping sub-datasets to learn the distinct morphology of each artifact. Labels are transformed to take the artifact class as a negative class. A fusion function integrates output from all experts to form a predictive probability distribution for the final prediction. A meta-learned probability threshold is applied to maximize the sensitivity of the MoE.}
    \label{fig:ensemble_learning}
\end{figure}

\subsection{Mixture of Experts}\label{sec:ensembles}
The "mixture of experts (MoE)" DL approach is often confused with deep ensembles. A deep ensemble combines DL models trained on the same data using different seed initializations or hyperparameters to learn different aspects of the data~\cite{zhang2021understanding}. Unlike deep ensemble, in MoE, each DL model is trained for a specific task (blur, fold, blood, folded tissue, and damaged tissue detection) to become a specialist in particular task of data. Instead of applying simple majority voting like deep ensembles, a gating mechanism forms the final prediction, incorporating output from diverse experts and improving robustness.     

Our proposed DL scheme is a kind of MoE where we integrate five identical DL architectures (also called base learners or experts) after training on the parts of the data (similar to bagging). Bagging offers the advantage of reducing variance, thus eliminating overfitting by training models on subsets of data. This parallel and data-independent training strategy avoids affecting the results of other experts. We form two MoE-based DL pipelines, ViTs-based MoE and DCNNs-based MoE, by choosing five base learners (DCNN or ViT architectures as explained in Sec.~\ref{sec:models}). All these experts are trained on five overlapping subsets, $\{\mathcal{D}_{blood}, \mathcal{D}_{blur}, \mathcal{D}_{airbubble}, \mathcal{D}_{damaged}, \mathcal{D}_{folds}\} \in \mathcal{D}$. Each sub-dataset contains a distinct artifact class and the same artifact-free class as shown in Figure~\ref{fig:ensemble_learning}. For simplicity, we transform ground truth labels as a positive class with the label `1' for artifact-free and a negative class with the label `0' for the artifact class. 

The contingent MoE model, $\Omega$, forms a single prediction using the aggregation function ($G$). $G$ is similar to gating, which combines the output probabilities of the experts using a fusion approach. In short, the proposed approach formulates MoE trained on individual artifact morphology detection tasks. For artifact models $\psi_i \in \{\psi_{blood}, \psi_{blur}, \psi_{airbubble}, \psi_{damage}, \psi_{fold}\}$, we only utilize the prediction for negative class ($P_{\psi_0}$) (a.k.a probability of being an artifact), and fuse binary outputs for $\Omega$ as shown in Eq.~\eqref{eq:ensemble_probs}. 

\begin{equation} 
    \boldsymbol{{P_\Omega}}_{y} = G(\psi_{blood}, \psi_{blur}, \psi_{airbubble}, \psi_{damage}, \psi_{fold)}
\end{equation}

\begin{equation}\label{eq:ensemble_probs}
    \centering 
    \boldsymbol{{P_\Omega}}_{y} = 
    \begin{cases}
    1 - max (P_{\psi_{i_0}}) \qquad \textrm{for artifact-free (positive) class} \\
     max (P_{\psi_{i_0}}) \qquad  \quad\quad \textrm{for artifact (negative) class}
   \end{cases}
\end{equation}

To evaluate the final prediction ($\boldsymbol{{\hat{P}_{\Omega_y}}}$), we adopt a form of meta-learning by placing a constraint on maximizing the sensitivity of the model for the positive (artifact-free) class. Therefore, we introduce a probability threshold, $t_s$, to handle previously unseen tissue morphology and avoid misclassifying artifact-free patches with potential diagnostic relevance. In other words, if the probability of being a positive class in $\boldsymbol{P_{\Omega_y}}$ is higher than $t_s$, then we assign \emph{artifact-free} label to the patch as shown in Eq.~\eqref{eq:final_ensemble_pred}. Here, $t_s$ would help to efficiently minimize false negatives without re-training models with a new cohort of WSIs with different tissue types or staining. We determine the best value of $t_s$ by maximizing the true positive rate (sensitivity) in the receiver operating characteristic (ROC) curve over the validation data.     

\begin{equation}\label{eq:final_ensemble_pred}
    \centering
    \boldsymbol{\hat{P}_{\Omega_y}} =
    \begin{cases}
     Artifact-free \qquad \textrm{if} \quad P_{\Omega_{y_0}} \geq t_s \\
    Artifact_k \qquad \qquad \textrm{Otherwise} \quad  k \in\{1,2,3,4,5\}
   \end{cases}
\end{equation}

\subsection{Multiclass Models}
In case of multiclass models ($\psi_{multi}$) with predicted probability distribution $\boldsymbol{P_{\psi_{y_i}}} \forall \quad i \in \{0,1,2,3,4,5\}$. 
We find the probability threshold ($t_s$) by maximizing sensitivity similar to MoE (see Section~\ref{sec:ensembles}). In other words, if the predicted probability for the artifact-free class is higher than $t_s$, then the patch is assigned \emph{artifact-free} label. Otherwise, the artifact label with the highest probability value is assigned (see Eq.~\eqref{eq:multiclass_pred}).

\begin{equation}\label{eq:multiclass_pred}
    \centering 
    \boldsymbol{\hat{P}_{\psi_{multi_y}}}  = 
  \begin{cases}
     Artifact-free \quad \textrm{with} \quad  p_{\psi_{y_0}} \qquad  \textrm{if} \quad p_{\psi_{y_0}} \geq t_s \\
    Artifact_k \quad \textrm{with}\quad  p_{\psi_{y_k}} \quad max(p_{\psi_{y_1}}, p_{\psi_{y_2}}, ..., p_{\psi_{y_k}}) \quad \textrm{Otherwise}
   \end{cases}
\end{equation}

\subsection{Post-processing}
At the inference stage, we utilize predictions for both artifact detection and QC applications, as illustrated in the post-processing part of Figure~\ref{main}. Since the predictions of DL models are patch-based, we need to stitch patches back to see the overall view of the tissue in the WSI structure. However, stitching smaller patches introduces boundary artifacts (blockish appearance)~\cite{khened2021generalized}. To avoid this problem, we turn to the matrix-filling approach. 

For patch $x_i$ with coordinates $(x_0,y_0)$, the next consecutive patch $x_{(i+1)}$ holds the difference of sampling stride ($s$) with coordinates $(x_1,y_1)=(x_{0+s},y_{0+s})$. Here, $s$ equals the patch size owing to a uniform, non-overlapping grid. For the segmentation map, we use a matrix ($M$), a downscale version of the original resolution, to assign predicted class $k$.

\begin{equation}
\centering
\begin{aligned}
    M[x_0:x_0+s, y_0:y_0+s] = k \qquad \textrm{where} \quad s=224 \quad \textrm{(patch-size)}\\
    M[x_1:x_1+s, y1:y_1+s] = k \qquad \textrm{where} \quad k=\{0,1,..5\}
\end{aligned}
\end{equation}

\begin{figure}[h]
    \centering
    \includegraphics[width=1.03\textwidth]{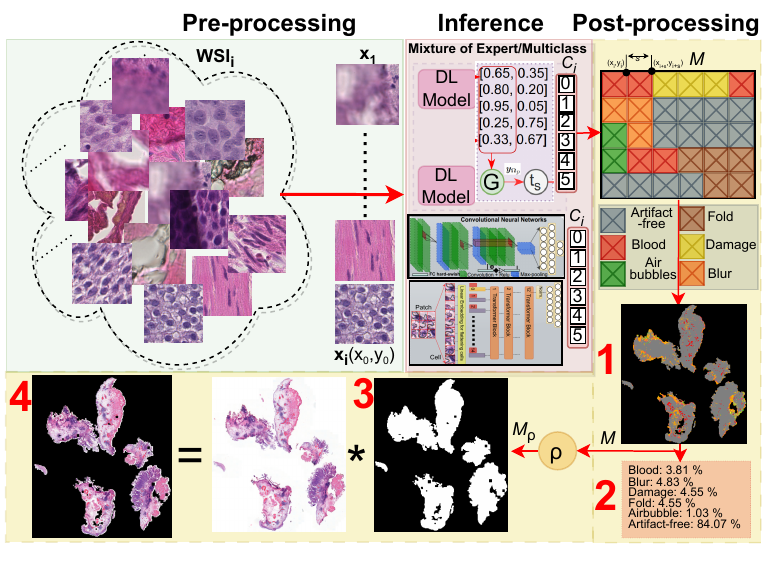}
    \caption{\textbf{Overview of deep learning pipeline emphasizing the post-processing stage during the inference.} \emph{Pre-processing:} The whole slide image (WSI) is split, and every patch is stored with its corresponding coordinate. \emph{Inference:} Every patch is assigned a label using a mixture of experts or multiclass DL models. \emph{Post-processing:} The matrix-based filling method assigns a color to every pixel (in the down\-scaled version of WSI) at the corresponding coordinate location. Post-processing provides: 1) Segmentation map; 2) Artifact report for quality control; 3) Artifact-free region of interest map, and 4) Artifact-refined WSI for computational analysis.}
    \label{fig:postprocess}
\end{figure}

Since $M$ is down-scaled to sampling stride size, every filled box can be seen as a pixel in the final segmentation map (see 1 in Figure~\ref{fig:postprocess}). We use filled-in $M$ for the artifact report to calculate the percentage of predicted patches with artifact class $k$ over the total patches $N_{tot}$ in the foreground. See 2 in Figure~\ref{fig:postprocess} for an example artifact report for QC. 

\begin{equation}
    Per_{k} = \frac{N_{k}}{N_{tot}} * 100\% \qquad \textrm{where} \quad N_{k}=\textrm{predicted with class}\quad k
\end{equation}

We denote the artifact-free post-processed region as $\rho$. It measures the usefulness of the WSI and can be compared against a predefined threshold $\tau$ for assessing its suitability (accepting or discarding) for developing DL algorithms.

\begin{equation}
    \rho = \frac{\textrm{Number of artifact-free pixels} \quad (N_{k_0})}{\textrm{Total number of pixels in the foreground}\quad (N_{tot})}
\end{equation}

To highlight the histologically relevant region, we binarize $M$ to $M_{\rho}$ and treat all artifact classes as a single class, as shown in Eq.~\eqref{eq:binary_mask}. The binary mask ($M_{\rho}$) indicates the potentially histologically relevant RoI (see 3 in Figure~\ref{fig:postprocess}). Later, we apply a morphological closing operation to remove small holes in the final mask. 

\begin{equation} \label{eq:binary_mask}
    M_{{\rho}_{(i,j)}} =
    \begin{cases}
        1, \qquad \textrm{if} \quad M_{(i,j)} = k_0 \quad \textrm{(artifact-free)}\\
        0, \qquad \textrm{Otherwise}
    \end{cases}
\end{equation}

Finally, obtain artifact-free WSI by performing the Hadamard product between $M_{\rho}$ and the original WSI ($I \in \mathbb{R}^{m \times n}$) with the dimensions of $m \times n$ (see Eq.~\eqref{eq:elementwise}). Using the nearest interpolation, we resize the $M_{\rho}$ mask to $m \times n$. Let us denote the element at the $i$-th row and $j$-th column of $M_{\rho}$ as $M_{\rho}(i, j)$, and the corresponding element in $I$ as $I(i, j)$. This element-wise operation between $M_{\rho}$ and $I$ removes any regions or areas with the presence of artifacts (see 4 in Figure~\ref{fig:postprocess}) and $I_{artifact-free}$ can be written as:

\begin{equation}\label{eq:elementwise}
\large
    (I\odot M_{\rho})_{ij} = 
    \begin{bmatrix} 
    M_{\rho_{(1,1)}} \cdot I_{_{(1,1)}} & M_{\rho_{(1,2)}} \cdot I_{_{(1,2)}}& \dots & M_{\rho_{(1,n)}} \cdot I_{_{(1,n)}} \\ M_{\rho_{(2,1)}} \cdot I_{_{(2,1)}} & M_{\rho_{(2,2)}} \cdot I_{_{(2,2)}} & \dots & M_{\rho_{(2,n)}} \cdot I_{_{(2,n)}} \\ \vdots & \vdots & \ddots & \vdots \\ M_{\rho_{(m,1)}} \cdot I_{_{(m,1)}} & M_{\rho_{(m,2)}} \cdot I_{_{(m,2)}} & \dots & M_{\rho_{(m,n)}} \cdot I_{_{(m,n)}} 
    \end{bmatrix}
\end{equation}

\subsection{Evaluation Metrics}
For performance comparison, we report accuracy, sensitivity, and the F1-score. Let TP, FN, FP, and TN denote true positive and false negative, false positive, and true negative predictions, respectively. Here, a positive class refers to a patch without artifacts (artifact-free patch). Then, the confusion matrix (CM) is a tabular representation of the model's predictions using TP, FN, FP, and TN. Accuracy is the proportion of correct predictions to the total number of predictions and is defined as $Acc. = (TP+TN)/(TP+FN+FP+TN) $. Sensitivity, also known as recall, measures the proportion of actual positives correctly identified by the model and is termed $Sens. = TP/(TP+FN) $. High sensitivity is essential to retaining potentially relevant (artifact-free) RoIs for the diagnostic algorithm. On the other hand, specificity $Specs. = TP/(TP+FP) $, quantifies the performance of a model in distinguishing negative instances from those falsely labeled as positive. In our application, high specificity filters out irrelevant information (artifacts) appearing in relevant RoIs. The F1 score is the harmonic mean of precision and recall and is calculated as $F1 = 2\cdot(\textrm{precision} \cdot \textrm{recall})(\textrm{precision} + \textrm{recall})$, where precision =  $TP/(TP+FP)$. For overall segmentation, dice co-efficient is reported. Dice scores the overlap between the predicted segmentation and the ground truth and ranges from 0 to 1, where 1 indicates perfect overlap between the predicted and ground truth segmentation. We use model weights with the lowest validation loss during the training to report these evaluation metrics. 

We have considered FLOPS, parameters, and inference time for computational complexity evaluation. FLOPS measures the number of floating-point operations required by a specific algorithm. The number of parameters refers to the learnable parameters in the model that are used to perform operations, where high parameters result in more FLOPS. Finally, inference time is the time the DL model consumes to make predictions over a patch. These metrics, combined, provide a comprehensive understanding of the DL model's performance and computational efficiency, which are crucial for assessing the practical applicability in real-world scenarios.

\subsection{Implementation Details} \label{sec:experimental_detail}
The code was implemented using Python. The patch extraction was accomplished using the Pyvips~\footnote{\burl{https://libvips.github.io/pyvips/}} library. During the patching, we used torch multiprocessing~\footnote{\burl{https://pytorch.org/docs/stable/multiprocessing.html}} to carry out process pooling for faster pre-processing.
The extracted patches were standardized to the mean and standard deviation of ImageNet~\cite{imagenet} due to transfer learning over ImageNet weights. To compensate for the scarcity of labeled data, augmentation is applied at each epoch during the training~\cite{morales2021artificial, wetzer2023representation}. We used random geometric transformations, including rotations and flips, both horizontally and vertically. Our DL models consist of a feature extractor and a classifier with three fully connected (FC) layers. We used state-of-the-art architectures MobileNetv3~\cite{mobilenet} and ViT-Tiny~\cite{Deit} as backbones for feature extractors.
MobileNetv3 was borrowed from the Pytorch~\footnote{\burl{https://github.com/pytorch/pytorch}} DL framework, and ViT-Tiny was taken from the Timm~\footnote{\burl{https://timm.fast.ai/}} library. Both of these backbones were initialized with ImageNet weights. We referred to the best hyperparameter settings from works~\cite{kanwal2022quantifying,kanwal2023vision,kanwal2024sure} and fixed final parameters to cross-entropy loss, SGD optimizer, ReduceLRonPlateau scheduler initialized with 0.01, batch size of 128, early stopping of 20 epoch over the validation loss to avoid overfitting, dropout of 0.2, and fixed random seed for reproducibility. 
All training and inference experiments were done on the Nvidia A100 40GB GPU. 
The source code is available at \href{https://github.com/NeelKanwal/Equipping-Computational-Pathology-Systems-with-Artifact-Processing-Pipeline}{Github}.

\section{Experimental Results and Discussion} \label{sec:results}
This section presents experimental results for training and validating DL pipelines of the EMC cohort and discusses their performance on validation, testing, and external data. 

\begin{table}[h]
\caption{\textbf{Performance of artifact processing pipelines on the validation set of \emph{EMC}$_{dev}$ cohort~\ref{dataset:emcdev}.} Various DL pipelines, including the mixture of experts (MoE) and multiclass models using SOTA DCNN and ViT architectures, are deployed. A simple binary formulation is used for a fair comparison, and accuracy for the artifact-free class is reported. The best results are marked in bold, and the second-best results are underlined in each column.}
    \centering
    \begin{tabular}{||l|l|| c|c|| c|c|c||}
    \hline
    \multicolumn{2}{||c||}{DL architecture} & $Acc. (\%)$ & $F1$ & $Acc._{afree}$ & $F1_{afree}$  & $Sens._{afree}$ \\ [0.2em]
    \hline
    {} & MoE & 92.08 & 91.87 & \underline{97.82} & \underline{88.66} & 90.12 \\[0.2em]
    {} & Multiclass  &  93.48 & 93.43 & 94.96 & 78.64 & \textbf{96.89} \\ [0.2em]
    \multirow{-3}{*}{\rotatebox{90}{DCNNs}} & Binary & \underline{95.92} & \underline{95.26} & - & - & \underline{94.68} \\[0.2em]
    \hline
    {} & MoE  & 94.81 & 94.53  & \textbf{97.84}  & \textbf{89.06} & 91.92 \\[0.2em]
    {} & Multiclass & 94.29 & 94.48 & 96.79 & 83.80 & 86.84 \\[0.2em]
    \multirow{-3}{*}{\rotatebox{90}{ViTs}}  & Binary & \textbf{97.45} & \textbf{97.46} & - & - & 87.25 \\[0.2em]
    \hline
    \end{tabular}
    \label{table:2}
\end{table}

\begin{table}[h]
\caption{\textbf{Comparison of the proposed mixture of experts (MoE) against the literature on identical classification tasks.} Note that the reported methods were developed using different data under different experimental setups. Thus, the results are provided for reference, not as a direct comparison. The best results are marked in bold, and the second-best results are underlined.}
    \centering
    \begin{tabular}{||l|c||}

    \hline
     Task and method from literature  & Accuracy (\%) \\ [0.2em]
    \hline
     Folded tissue detection by \cite{Shakhawat2020} & 92.17 \\
     Folded tissue detection by \cite{babaie2019deep} & 96.7 \\
     Blur detection by \cite{senaras2018deepfocus} & 93.2\\
     Blur detection by \cite{albuquerque2021deep} & 94.4 \\
     Air bubble detection by \cite{Shakhawat2020} & 87.33 \\
     Air bubble detection by \cite{raipuria2022stress} & 91.5 \\
     Blood detection by \cite{swiderska2016automatic} & 85 \\
     Damaged tissue detection by \cite{swiderska2018deep} & 90 \\
     Five artifacts - MoE-DCNNs (Ours) & \underline{97.82} \\
     Five artifacts - MoE-ViTs (Ours) & \textbf{97.84} \\
    \hline\hline
    \end{tabular}
    \label{table:2_sota}
\end{table}

\subsection{Validation on the \emph{EMC}$_{dev}$ Cohort}
This experiment aims to evaluate the performance of the proposed MoE and multiclass models for artifact detection task. These pipelines consist of four DL approaches using MoE and multiclass models based on DCNNs (MobileNetv3~\cite{mobilenet}) and ViTs (ViT-Tiny~\cite{Deit}). For simplicity, we will refer to DCNNs or ViTs in the discussion. For a baseline comparison, we also trained binary classification models (DCNN and ViT) using the entire \emph{EMC}$_{dev}$ dataset in a binary fashion. In other words, we wanted to compare the benefits and drawbacks of the simpler classification model against a MoE and their computational and performance trade-offs for efficient DL pipelines. 

We will first focus on discussing the performance aspect. Table~\ref{table:2} presents classification results over the \emph{EMC}$_{dev}$ validation subset. We have reported metrics for artifact-free classes to compare them fairly against baseline (binary) models. For better classification performance, we desire high sensitivity to avoid misclassifying artifact-free patches as artifacts and retain potential histologically relevant tissue for automated diagnostics. This is because the artifact detection application is not affected by one artifact class being classified as another. In the end, patches with the presence of any artifacts will be excluded from downstream (diagnostic) applications. Though the baseline models yield the best overall accuracy, they relatively underperform and exhibit lower sensitivity in classifying the artifact-free class.
The MoEs outperform multiclass models and baseline models in detecting artifact-free class. Overall, both MoE pipelines give superior results for the positive class and avoid false negatives. However, the DCNN-based multiclass model yields the best sensitivity score. Table~\ref{table:2_sota} shows validation results from other relevant works from the literature as a reference. The reported results can not be directly compared as the methods were trained using different data and varying experimental setup. To present an unbiased view, we test MoEs and multiclass models on unseen data from the same \emph{EMC}$_{dev}$ cohort.  

\begin{table}[h!]
\caption{\textbf{Generalization results on the test set of \emph{EMC}$_{dev}$ cohort~\ref{dataset:emcdev}.} The table presents results over unseen data, with and without probabilistic thresholding. All metrics are calculated for the classification performance over artifact-free class. The best results in each column are marked in bold, and the second-best results are underlined.}
    \centering
    \begin{tabular}{||l|l|| c|c|c||c|| c|c||}
    \hline
    \multicolumn{2}{||c||}{DL architecture}  & \multicolumn{3}{|c||}{Without probabilistic threshold} & {} & \multicolumn{2}{|c||}{With probabilistic threshold} \\
    \multicolumn{2}{||c||}{} &  Acc. (\%) & F1  & Sens. & \large{$t_s$} &  F1  & Sens. \\
    \hline
    &  MoE  & \textbf{97.82} & \underline{88.66} & 89.12 & 0.326 & \textbf{86.15} & \textbf{97.93} \\
         \multirow{-2}{*}{\STAB{\rotatebox[origin=c]{0}{{DCNNs}}}}  & Multiclass & 93.58 & 85.21 & \textbf{94.72} & 0.341 & 83.53 & 95.47 \\
        \hline
         & MoE & \underline{95.61} & \textbf{88.91} & \underline{90.45} & 0.052 &  \underline{84.90} & \underline{97.83} \\
        \multirow{-2}{*}{\STAB{\rotatebox[origin=c]{0}{{ViTs}}}}  & Multiclass  & 92.55 & 82.51 & 89.94 & 0.015 & 70.15 & 96.54  \\
         \hline
    \end{tabular}
    \label{table:3}
\end{table}

\begin{figure}[hb!]
    \centering
    \includegraphics[width= 0.9\textwidth]{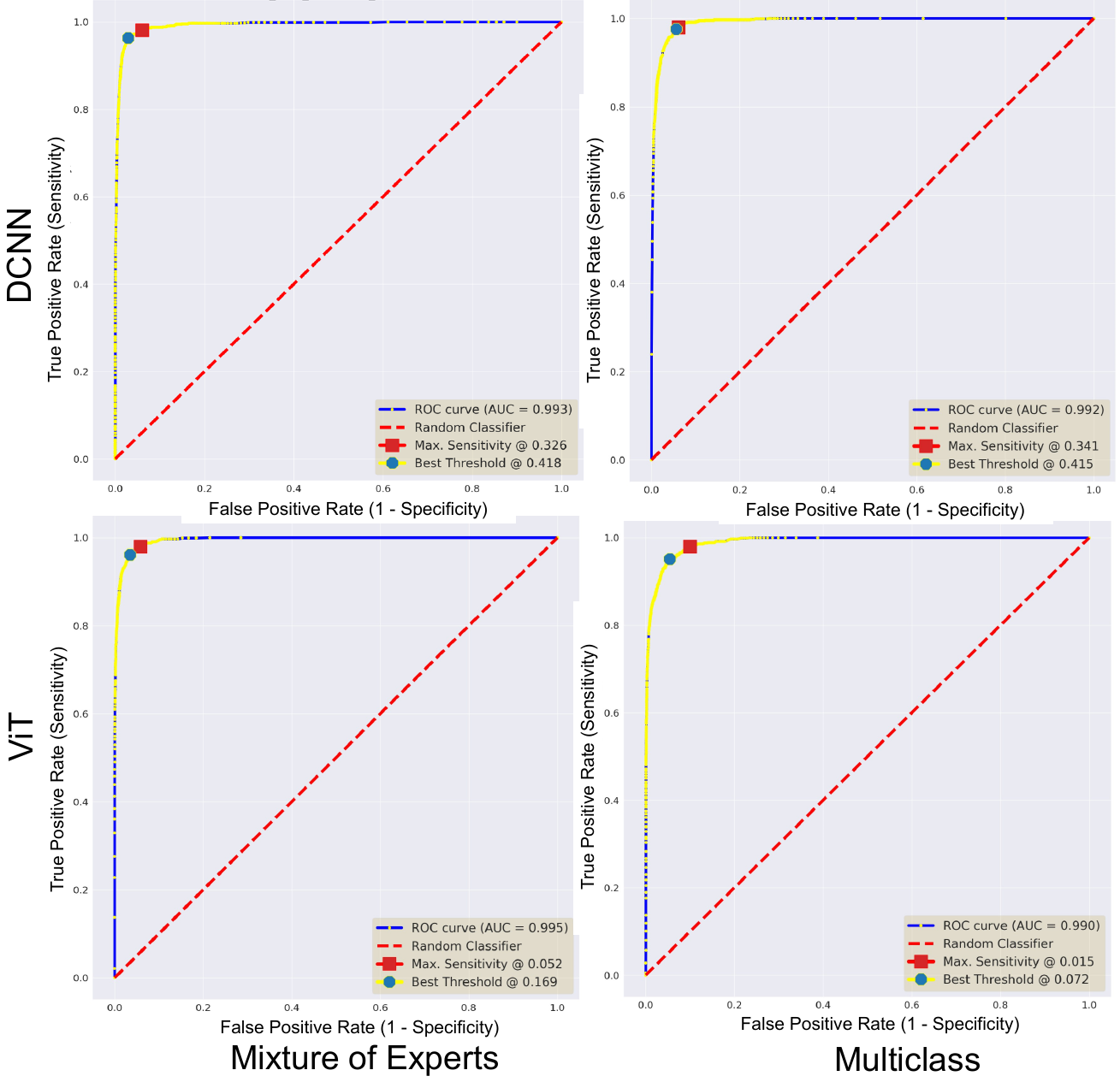}
    \caption{\textbf{ROC curves for deep learning pipelines over the validation subset.} All plots highlight the area under the curves (AUC) score and best probability thresholds for maximizing F1 and sensitivity metrics. }
    \label{fig:roc_curves}
\end{figure}

\begin{figure}[h!]
    \centering
    \includegraphics[width= 0.9\textwidth]{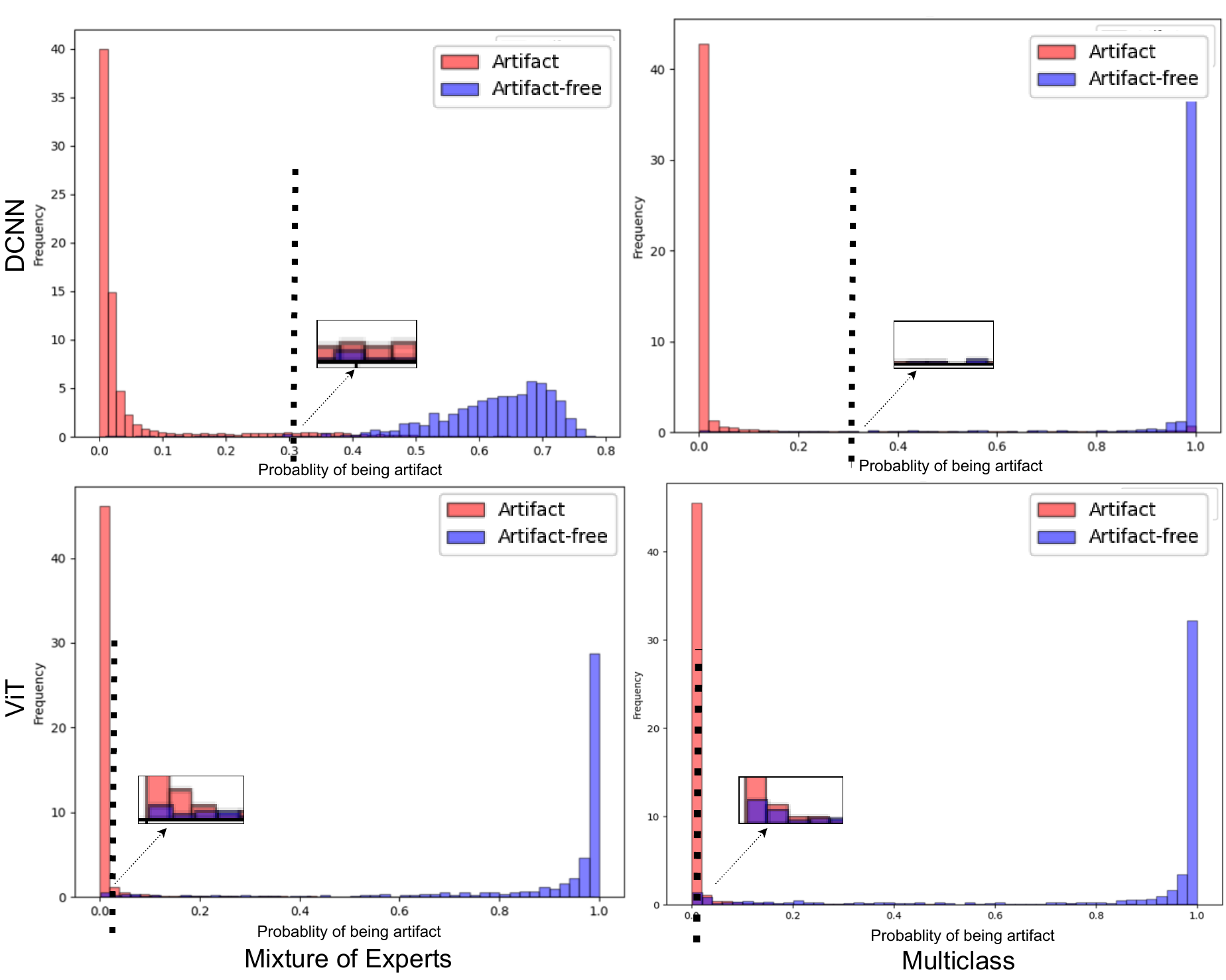}
    \caption{\textbf{Classification plots for deep learning pipelines over the validation subset.} All subplots highlight the delineation (black dotted line) with the estimated value of $t_s$ for probabilistic thresholding.}
    \label{fig:threshold_plots}
\end{figure}

We present generalization results in Table~\ref{table:3}. The table reports mixed results when probabilistic thresholding is not applied. To improve the sensitivity over new data, we learn a probability threshold ($t_s$) using ROC curves of the validation set (see Section~\ref{sec:ensembles}), as displayed in Figure~\ref{fig:roc_curves}. We target a 98\% sensitivity and obtain different $t_s$ values for each DL pipeline, as reported in Table~\ref{table:3}. Interestingly, the DCNN-based pipelines assign higher probability scores to the artifact-free class, indicating better confidence and stronger learning of histologically relevant morphology than the ViT-based models. Figure~\ref{fig:threshold_plots} reflects similar insight that ViT-based pipelines carry weak differentiation between artifacts and artifact-free patches (see black dotted line).  It is fascinating to see that probabilistic thresholding significantly improves the ability to detect artifact-free class, hinting that the proposed MoEs would be the best choice with the fewest false negatives. 

To evaluate the computational aspect, Table~\ref{table:4} indicates the computational complexity of all four DL pipelines. Undoubtedly, MoEs have nearly five times more parameters and lower throughput than multiclass models. This is because each MoE combines five binary experts. Comparatively, DCNN-based pipelines can be efficient at the inference stage due to very little patch processing time per second. We have to make a trade-off in selection, either choosing multiclass DCNN with better computational efficiency but relatively lower performance or based on the best performance. We prioritize classification performance and opt for the two best-performing DL pipelines from Table~\ref{table:3}; therefore, we will use MoEs for the following experiments.  

\begin{table}[h!]
\caption{\textbf{A comparative analysis of computational complexity.} Lower values of parameters and flops indicate computationally efficient models, and higher throughput is desired for faster inference.}
    \centering
    \begin{tabular}{||l|| c|c|c|c||}
            \hline
             DL Pipelines & \shortstack{Parameters\\ (M)$\big\Downarrow$} & \shortstack{Flops\\ (B)$\big\Downarrow$} & \shortstack{ Throughput\\ (p/sec.)$\big\Uparrow$}   \\
             \hline
            MoE (DCNNs) & 17.65 & 1.13 & 178  \\
            MoE (ViTs) & 27.62 & 5.38 & 128 \\
            Multiclass (DCNN) & 3.53 & 0.22 & 832 \\
            Multiclass (ViT) & 5.53 & 1.08 & 419  \\
            \hline
        \end{tabular}
        \label{table:4}
\end{table}

\begin{table}[ht!]
    \centering
      \caption{\textbf{Results for quantitative evaluation for assessing the robustness of the proposed mixture of experts (MoE) approach.} Qualitative evaluation is performed on external (out-of-distribution) data. The table reports classification performance corresponding to patch-wise classification and dice scores for overall segmentation maps obtained through artifact processing pipelines. }
    \label{table:5}
    \begin{tabular}{||c|l|c|| c| c|c|c||}
    \hline
         DL Pipeline & {Cohort} & {WSIs} & $F1_{afree}$  & $Sens._{afree}$ & $Spec._{afree}$ & Dice \\
    \hline
         & {} & $s_1$ & 92.86 & 93.48 & 53.76 & 0.909 \\
         & \multirow{-2}{*}{EMC$_{inf}$} & $s_2$ & 89.11 & 89.61 & 52.71 & 0.784 \\
         \cmidrule(lr){2-7}
         & & $s_3$ & 70.91 & 55.07 & 99.09 & 0.487 \\
         & \multirow{-2}{*}{SUH$_{inf}$} & $s_4$ & 85.51 & 79.78 & 44.57 & 0.572  \\
         \cmidrule(lr){2-7}
         & & $s_5$ & 60.05 & 43.99 & 80.53 & 0.532   \\
        \multirow{-7}{*}{\STAB{\rotatebox[origin=c]{90}{{\shortstack{MoE of DCNNs }}}}} & \multirow{-2}{*}{INCLIVA$_{inf}$} & $s_6$ & 37.39 & 23.55 & 98.97 & 0.506 \\
    \hline
         & & $s_1$ & 93.17 & 93.01 & 60.92 & 0.939  \\
         & \multirow{-2}{*}{EMC$_{inf}$} & $s_2$ & 89.34 & 87.97 & 63.18 & 0.795  \\
         \cmidrule(lr){2-7}
         & & $s_3$ & 68.79 & 54.51 & 79.56 & 0.367  \\
         & \multirow{-2}{*}{SUH$_{inf}$}  & $s_4$ & 87.97 & 85.63 & 26.38 &  0.482 \\
         \cmidrule(lr){2-7}
         & & $s_5$ & 78.92 & 66.02 & 79.71 & 0.559 \\
        \multirow{-7}{*}{\STAB{\rotatebox[origin=c]{90}{{\shortstack{MoE of ViTs}}}}} & \multirow{-2}{*}{INCLIVA$_{inf}$} & $s_6$ & 45.49 & 42.49 & 42.91 & 0.412 \\
        \hline \hline
    \end{tabular}
\end{table}

\subsection{Quantitative Evaluation}
\begin{figure}[h!]
    \centering
    \includegraphics[width= 1\textwidth]{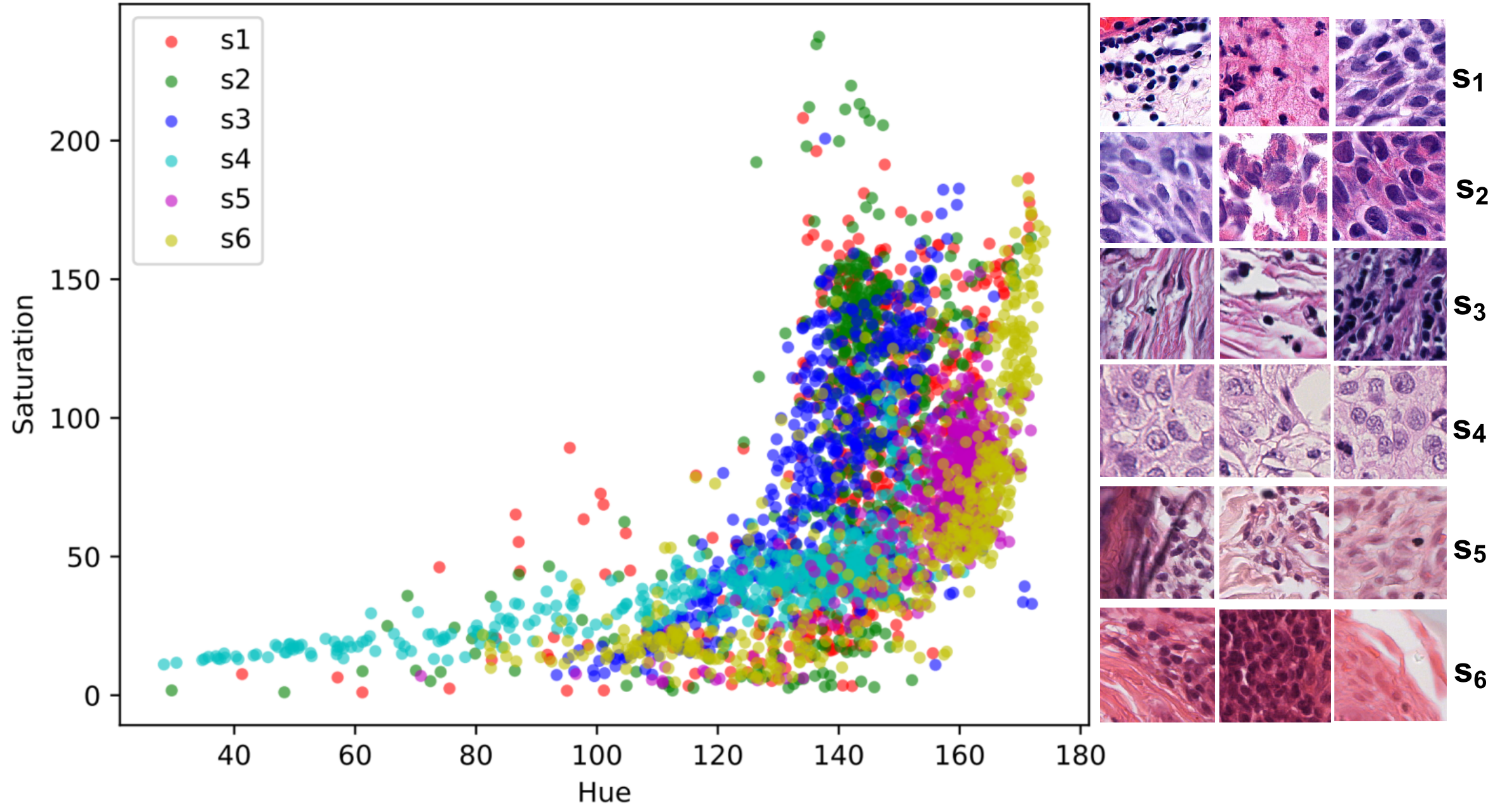}
    \caption{\textbf{Hue-Saturation plot shows massive variation in the external (out-of-distribution) data.} Random patches from all six WSIs ($s_1$-$s_6$) are used to calculate hue and saturation values to observe the depth of H\&E staining. WSI acquisition procedures from different laboratories and scanning hardware affect the final appearance of histological images (as shown on the right).\vspace{-2em}}
    \label{fig:variation}
\end{figure}

We perform this experiment to assess the robustness of DL pipelines over external (OoD) data. For this purpose, we chose six WSIs ($s_1$-$s_6$) from external validation data (see Section~\ref{sec:external}). Note that all these WSIs were prepared and scanned by different laboratories and scanning hardware. Thus, they exhibit vast differences in staining, tissue types, and image acquisition protocols, as displayed in Figure~\ref{fig:variation}. We did not incorporate color normalization in the artifact processing pipeline due to their additive computational cost and latency~\cite{kanwal2022quantifying}.

Quantitative assessment is crucial to objectively evaluate the numerical performance, enabling us to compare both the proposed MoEs of DCNNs and ViTs. We require histological correctness that only an expert can provide in the form of ground truths. Therefore, all WSIs were roughly annotated by FK, UK, and AM for different artifacts. Table~\ref{table:5} presents the results for classification and segmentation performance. Since certain artifacts, such as folded tissue, have blurry areas surrounded~\cite{kanwal2022devil}; one artifact class is likely to be predicted as another. Thus, for simplicity purposes, we report metrics for artifact-free (positive) classes only. 

\begin{figure}[ht!]
    \centering
    \includegraphics[width= 1\textwidth]{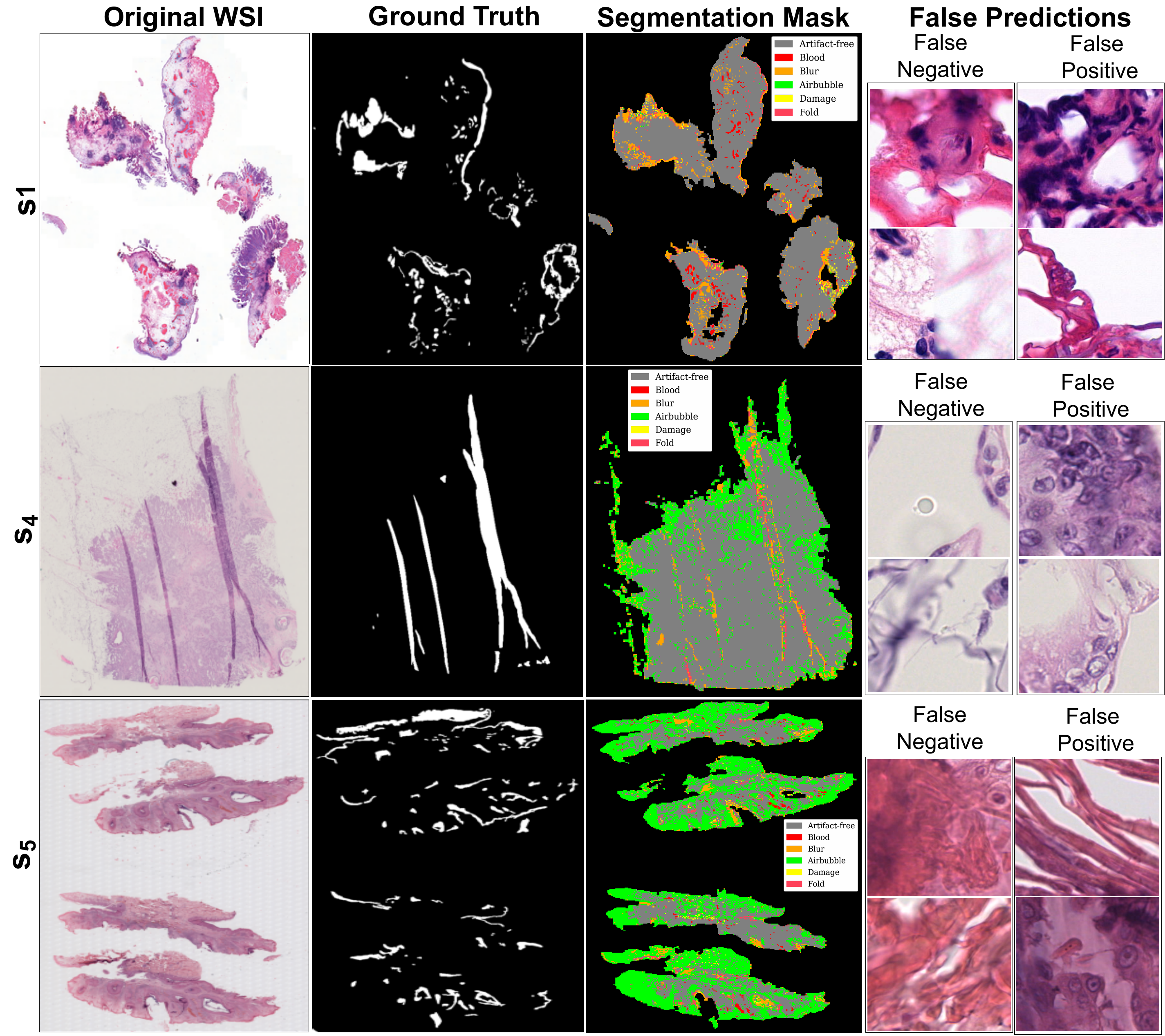}
    \caption{\textbf{Visualization of DCNNs-based mixture of experts' predictions with better performance over out-of-distribution data.} The image shows the original WSIs ($s_1$, $s_4$, and $s_5$) along with ground truth for artifacts (combined), artifact segmentation map, and a few examples of false predictions. False negative refers to patches detected as artifacts but were artifact-free, and false positive refers to patches detected as artifact-free but belonged to any artifact class.}
    \label{fig:result_cnns2}
\end{figure}

\begin{figure}[ht!]
    \centering
    \includegraphics[width= 1\textwidth]{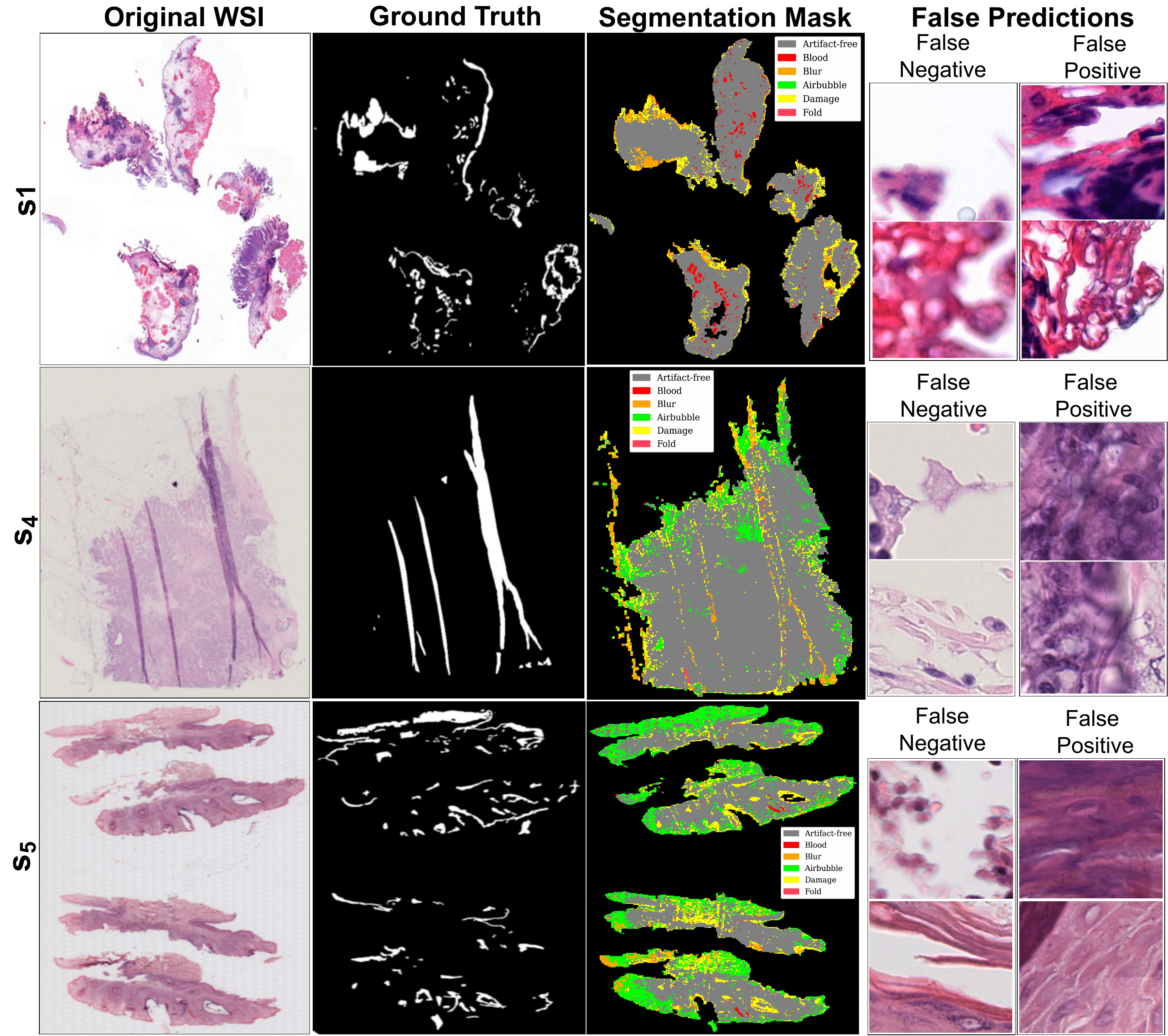}
    \caption{\textbf{Visualization of ViTs-based mixture of experts' predictions with better performance over out-of-distribution data.} Image shows original WSIs ($s_2$, $s_3$, and $s_6$) along with ground truth for artifacts (combined), artifact segmentation map, and a few examples of false predictions. False negative refers to patches detected as artifacts but were artifact-free, and false positive refers to patches detected as artifact-free but belonged to any artifact class.}
    \label{fig:result_vits2}
\end{figure}

\begin{figure}[ht!]
    \centering
    \includegraphics[width= 1\textwidth]{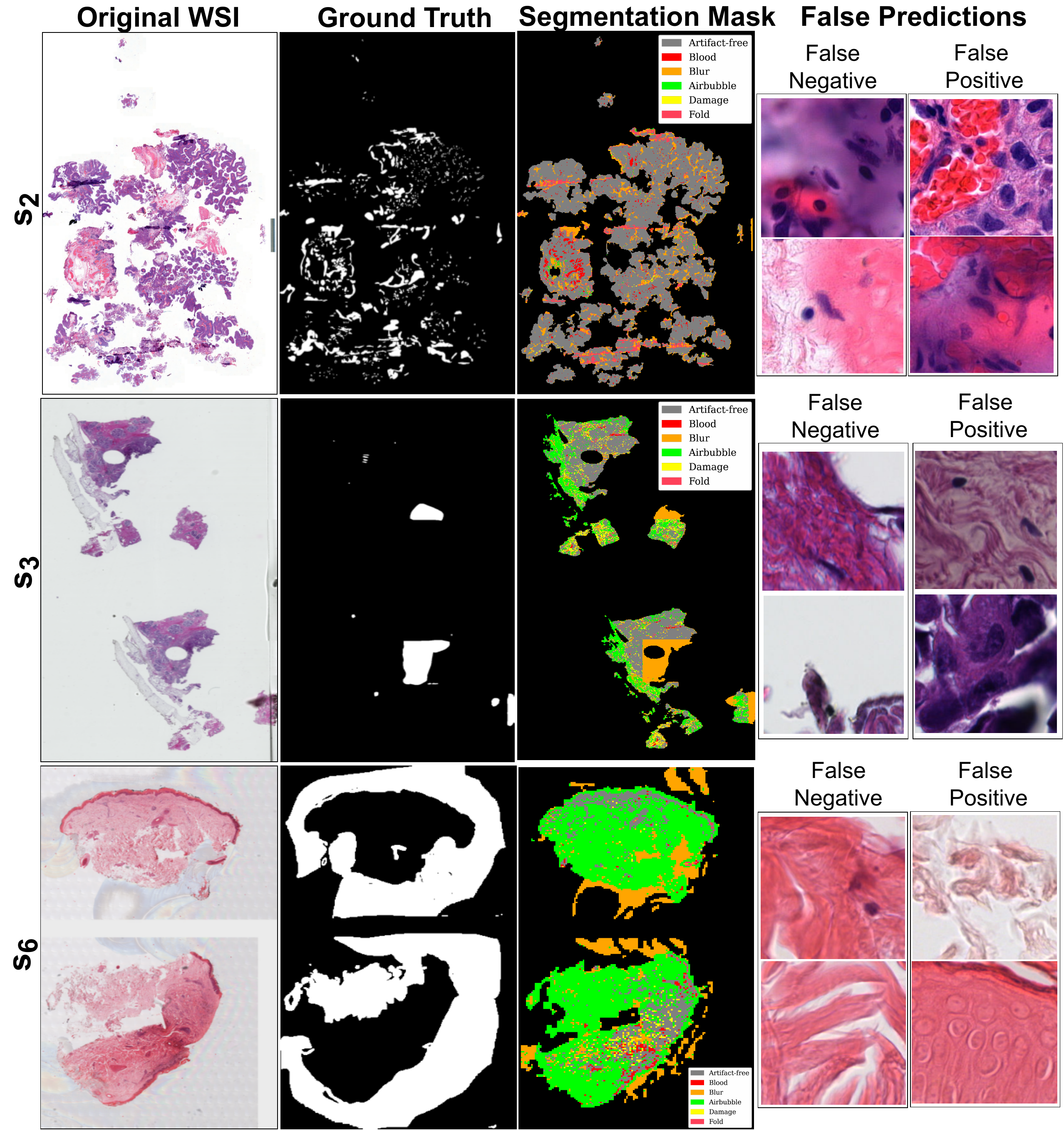}
    \caption{\textbf{Visualization of DCNNs-based mixture of experts' predictions with worst performance over out-of-distribution data.} Image shows original WSIs ($s_2$, $s_3$, and $s_6$) along with ground truth for artifacts (combined), artifact segmentation map, and a few examples of false predictions. False negative refers to patches detected as artifacts but were artifact-free, and false positive refers to patches detected as artifact-free but belonged to any artifact class.\vspace{-1em}}
    \label{fig:result_cnns}
\end{figure}

Both MoE pipelines experience a drop in sensitivity over breast cancer (\emph{SUH}$_{inf}$) and skin cancer (\emph{INCLIVA}$_{inf}$) WSIs. This behavior could be due to misclassifying ambiguous regions or susceptibility to specific tissue types. Since \emph{SUH}$_{inf}$ and \emph{INCLIVA}$_{inf}$ WSIs are OoD data for our DL pipelines, it is interesting to see that we get high specificity scores. In short, both pipelines ensure that most of the actual artifacts present in the data are accurately flagged. Dice score in Table~\ref{table:5} shows good segmentation results on the \emph{EMC}$_{inf}$ cohort. Nevertheless, \emph{EMC}$_{inf}$ is bladder cancer tissue and may carry more similarity in structural appearance. 

\begin{figure}[h!]
    \centering
    \includegraphics[width= 1\textwidth]{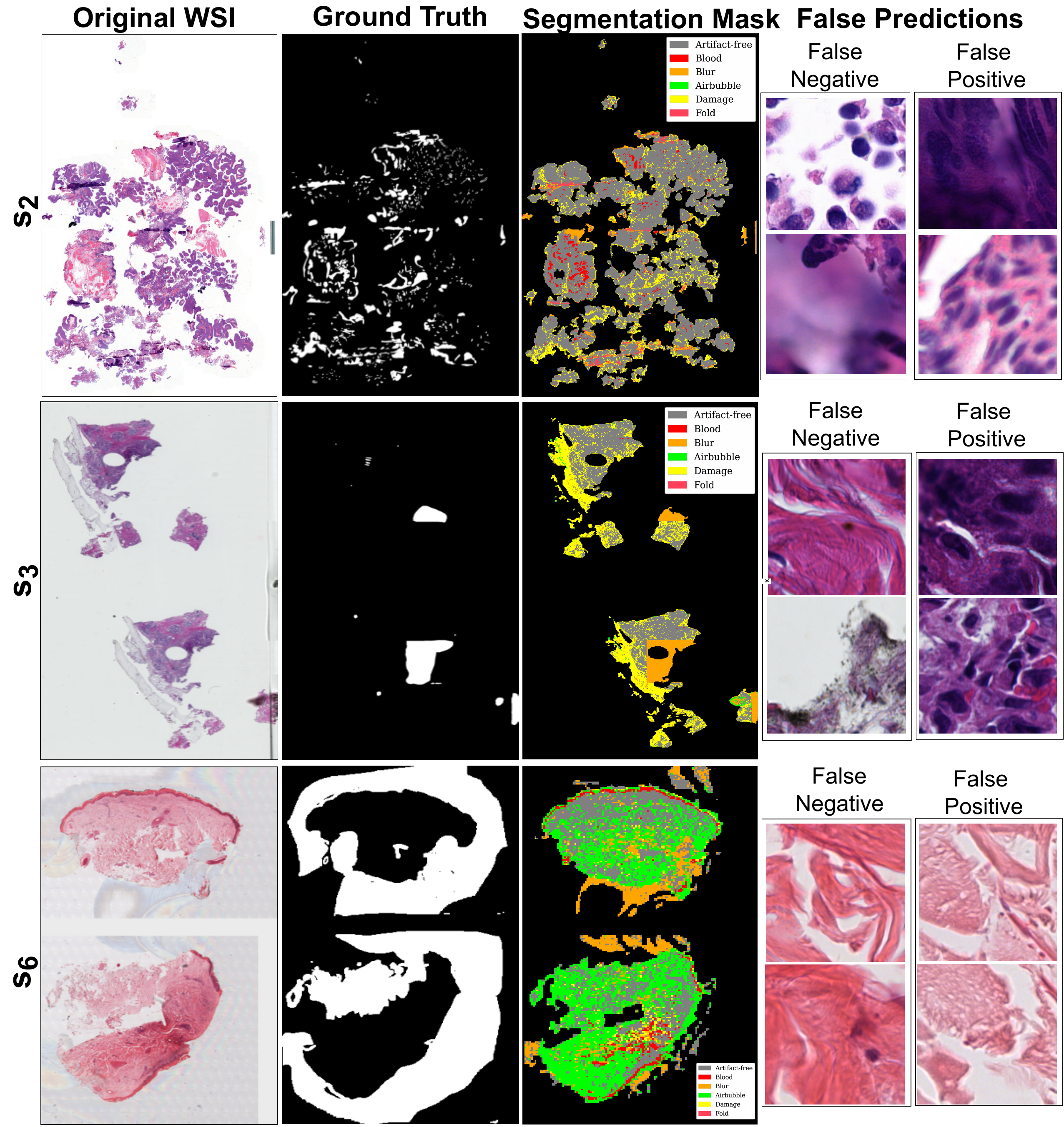}
    \caption{\textbf{Visualization of ViTs-based mixture of experts' predictions with worst performance over out-of-distribution data.} Image shows original WSIs ($s_2$, $s_3$, and $s_6$) along with ground truth for artifacts (combined), artifact segmentation map, and a few examples of false predictions. False negative refers to patches detected as artifacts but were artifact-free, and false positive refers to patches detected as artifact-free but belonged to any artifact class.}
    \label{fig:result_vits}
\end{figure}

Quantitative metrics can miss subtle nuances masked by overall performance scores. Therefore, we observe false predictions of both DCNNs-based MoE and ViT-based MoE over the better-performing cases ($s_1$, $s_4$, and $s_5$) and the worse-performing cases ($s_2$, $s_3$, and $s_6$) in OoD data. Figures~\ref{fig:result_cnns2} and ~\ref{fig:result_vits2} show ground truths and predictions masks for the better results in each cohort, and Figures~\ref{fig:result_cnns} and~\ref{fig:result_vits} show the same for the worst results in each cohort. Both MoEs densely predict artifacts in all three examples. Here, false negative instances pertain to regions identified as artifacts but were artefact-free. Conversely, false positives are cases classified as artifact-free but were labeled as any artifact class. Figure~\ref{fig:result_cnns} highlights that DCNN-based MoE might be overdoing their job predicting certain artifacts like air bubbles. For instance, in $s_6$, the entire WSI has a hazy appearance, with air trapped under most of the tissue. The false predictions for $s_6$ show that those examples lack cellular features. Likewise, for false positives in the case of $s_2$, those specific examples were the boundary of another artifact region and contained some presence of blood. In cases $s_2$ and $s_3$, annotations had some noise, and with the chosen mask overlap, the obtained ground truth was not accurate enough. On the other hand, the ViT-based MoE (in Figure~\ref{fig:result_vits}) appears to be slightly overdoing damage detection. In most false predictions here, we might be dealing with potentially noisy and imprecise ground truth annotations. Therefore, relying on only quantitative analysis is not concrete and conclusive. We require a thorough qualitative analysis by field experts to scrutinize further the strengths and weaknesses of both MoEs in detecting artifacts.

\begin{figure}[htb!]
    \centering
    \includegraphics[width= 0.8\textwidth]{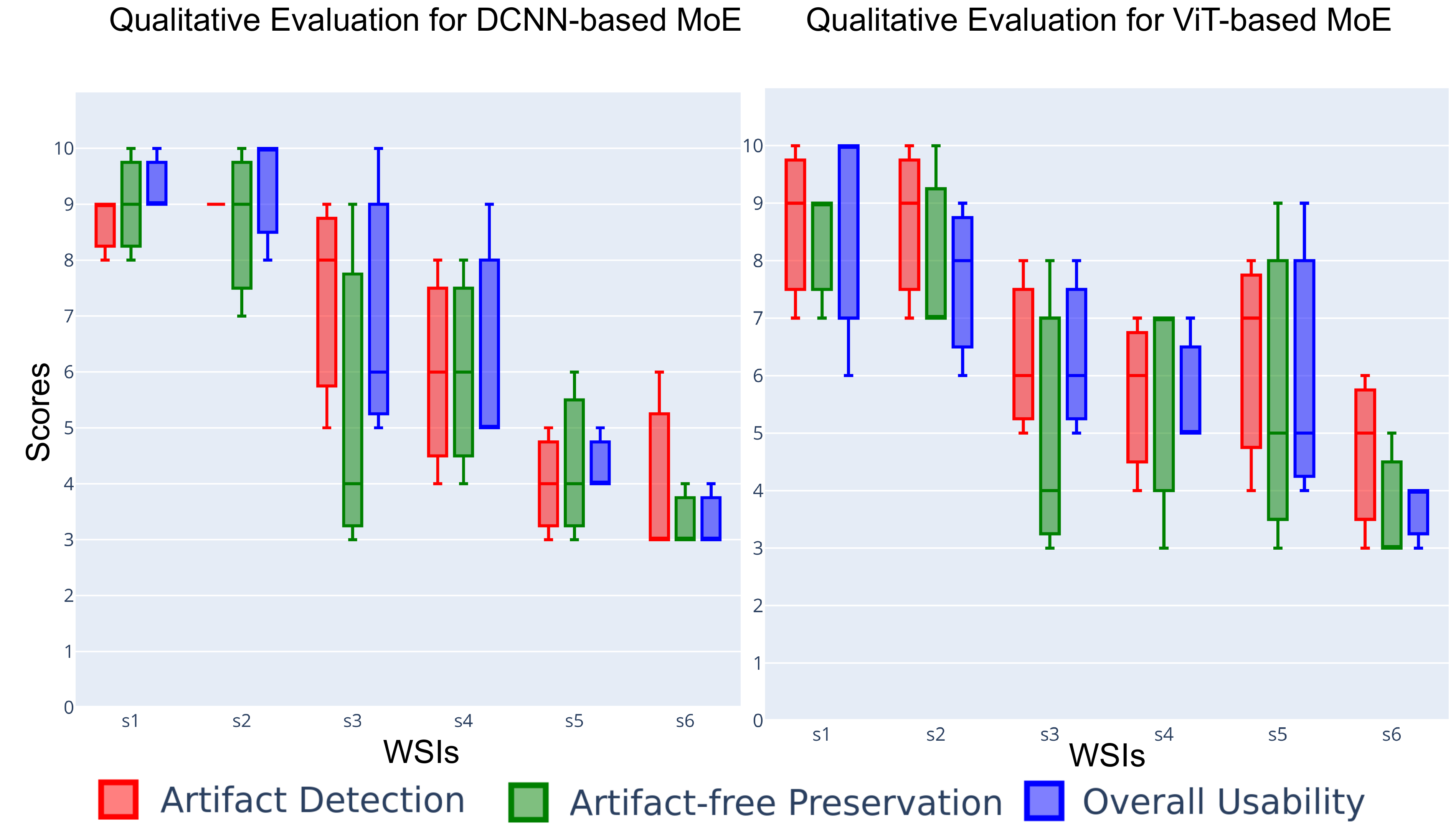}
    \caption{\textbf{Scores for qualitative evaluation by field experts (P1, P2 and P3) for different Tasks.} The boxplot provides a visual representation of the experts' assessments for predictions of OoD WSIs. The scores were provided on a scale of 1 to 10, with higher scores indicating better performance. }
    \label{fig:boxplot}
\end{figure}

\begin{figure}[htb!]
    \centering
    \includegraphics[width= 0.85\textwidth]{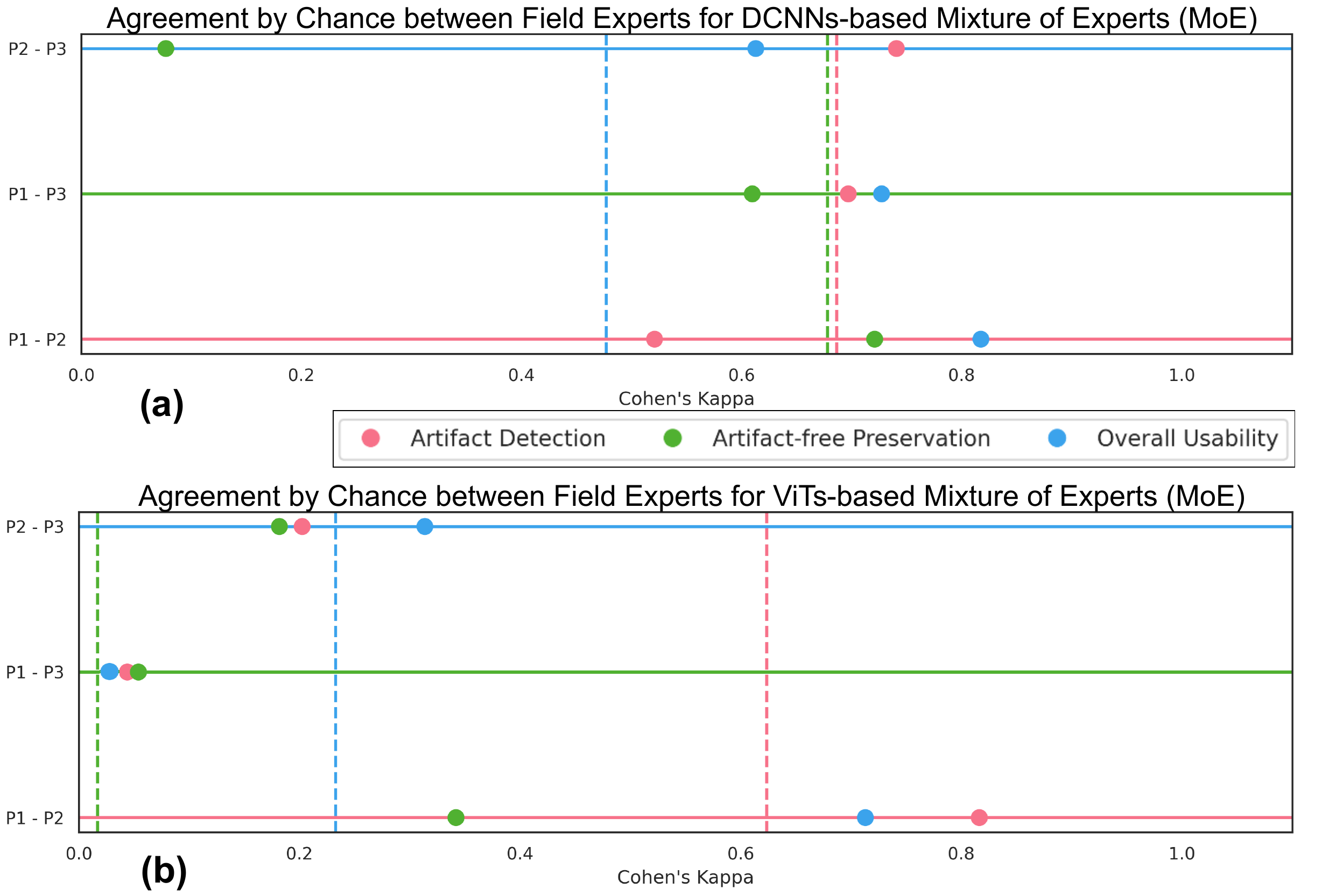}
    \caption{\textbf{Qualitative evaluation of artifact detection by the mixture of experts (MoE) models over OoD data.} Plots (a) represent Cohen's kappa score (on the x-axis) for DCNNs-based MoE and a pair of field experts on the y-axis, and (b) show scores for ViTs-based MoE. Both subplots show agreement by chance for each task. Each pair's average agreement of all three tasks is plotted as a vertical dashed line.}
    \label{fig:qualitative}
\end{figure}

\subsection{Qualitative Evaluation}
In this experiment, we perform qualitative evaluations by three field experts to delve deeper into the DL pipelines' behavior and see the holistic view after the artifacts refinement. While quantitative metrics provide valuable numerical insights into a model's performance, they often fall short of capturing the intricacies of segmentation results. Therefore, assessing whether the model was misclassified due to genuine limitations or imperfections in the ground truth is vital.

Three field experts (P1, P2, and P3) assessed segmentation maps for six WSIs ($s_1$-$s_6$) from three cohorts used in the above experiment. They scored them based on visual interpretation, including how well artifacts were detected, how artifact-free regions were preserved, and the overall diagnostic usability of WSIs after the artifact processing, where field experts scored them from 1 (worst) to 10 (best). Each expert who rated these WSIs was a domain specialist on a specific cancer type (see box plot in Figure~\ref{fig:boxplot}). Figure~\ref{fig:boxplot} represents the score variability for each task across the six WSIs. The central line in each box represents the median, while the box's upper and lower edges correspond to the interquartile range.

Cohen's Kappa coefficient measures the agreement between experts, where `1' indicates perfect agreement between experts and `0' indicates agreement no better than chance. Figure~\ref{fig:qualitative} reveals levels of agreement for each assessment category among the different pairs of experts for DCNNs-based MoE and ViT-based MoE. Vertical dotted lines present the average consensus across three assessment categories for each pair (in corresponding color). Both subplots highlight substantial agreement for overall usability and high average agreement between P1 and P2 (red dashed line) in Figure~\ref{fig:qualitative}. In contrast, artifact-free preservation has relatively lower agreement, echoing similar findings across all pairs. Based on the remarks obtained from field experts (see Figure~\ref{fig:boxplot}), generally, better results were obtained for bladder cancer WSIs ($s_1$, $s_2$). Although MoEs were too sensitive for detecting blurry areas, their folded and damaged regions were well segmented. In breast cancer WSIs ($s_3$, $s_4$), adipose tissue was predicted as air bubbles (with DCNNs-based MoE) or damaged (with ViTs-based MoE). Note that the training data did not include adipose tissue, primarily fat cells. This situation can be more evident in breast samples because there is more adipose tissue in them than in other cancer types. While adipose tissue can provide valuable contextual information and aid in certain aspects of diagnosis, its absence does not necessarily preclude an accurate assessment of breast cancer. In practical scenarios, there can also be other artifacts, like pen markings, due to manual annotation of RoIs. Since the models are trained for artifact-free class, we may expect MoE to distinguish between artifact-free and other regions. It is due to the fact that pen marking gives a similar visual of folded or blurry class patches as the cellular feature or tissue texture diminishes and becomes hard to observe. Nevertheless, encountering patches with pen markings might disappear as DP workflow becomes standard in all labs, and annotation over glass slides will be done manually rather than with markers. 

The particular examples of skin cancer WSIs ($s_5$, $s_6$) had significant air bubbles, leaving a hazy and unclear appearance over the foreground tissue. At the same time, both artifact processing pipelines were overdoing air bubble prediction, and the epidermis was predicted as blood. The performance of both MoEs is worst in these cases; one of the reasons could be the severity of artifacts and significant variation in staining in the WSI. While there is generally substantial agreement among field experts for overall diagnostic usability, there are areas, such as artifact-free preservation, where discrepancies emerge and may be more challenging to achieve. Moreover, considering inter-rater variability, DCNNs-based MoE indicates potential effectiveness for artifact detection and overall diagnostic usability.

By triangulating quantitative and qualitative analysis findings, we conclude that DCNNs-based MoE provides better generalizability and robustness with the trade-off of higher computational cost.

\section{Conclusion} \label{sec:conclusion}
In this work, we established end-to-end deep learning (DL) pipelines, taking whole slide images (WSIs) as input and providing artifact-refined WSIs to enable computational pathology (CPATH) systems to make reliable predictions. For the development of DL pipelines, we propose the mixture of experts (MoE) scheme and multiclass models. The MoE scheme uses five base learners (experts) with underlying state-of-the-art DL architectures (MobileNetv3 or ViT-Tiny). The MoE captures the intricacies of different artifact morphologies and dynamically combines predictions using a fusion mechanism to generate predictive probability distribution. Later, a meta-learned probabilistic threshold is applied to improve sensitivity for histologically relevant regions. In rigorous experiments, we performed generalizability and robustness tests over DL pipelines by testing on external cohorts of different tissue types. During the investigation, we found that the MoE scheme with underlying DCNNs attains the best classification and segmentation performance with some computational trade-offs compared to multiclass models. However, if high inference speed is the desired requirement, then multiclass models are a better choice with some degree of performance trade-off. Furthermore, during the qualitative evaluation, field experts rated the outcomes and agreed substantially on the overall usability of DCNNs-based MoE.    

Our artifact-processing DL pipelines can provide various outcomes, such as a segmentation map, artifact report, artifact-free mask with potential regions of interest with histological relevance, and an artifact-refined WSI for further computational analysis. Overall, the proposed DL solution is efficient and has a significant advantage in equipping the CPATH system with the necessary tools to isolate anomalies (or noise) from affecting automated clinical applications.

\section{\textcolor{blue}{Limitations and Future Work}}\label{sec:limitations}
The proposed work has a limitation in that the DL models were trained on a dataset prepared from a single cohort of data. In future work, these limitations can be overcome by pooling datasets from different cohorts in training and adopting an active learning strategy to adjust meta-learned thresholding parameters for improved sensitivity. Also, by formulating tailored fusion mechanisms for different cancer types. Furthermore, artifact-refined WSIs can be tested with the corresponding diagnostic or prognostic algorithms to assess the usefulness of artifact processing pipelines for clinical practice.


\section*{Abbreviations}
\begin{table}[h!]
    \centering
    \begin{tabular}{l l}
        WSI & Whole slide image \\
        DL  & Deep learning \\
        CPATH & Computational pathology \\
        MoE & Mixture of experts \\
        SOTA & state-of-the-art \\
        DCNN & Deep convolutional neural networks \\
        ViT & Vision transformer \\
        OoD & Out-of-distribution \\
        DP & Digital pathology \\
        QC & Quality control \\
        RGB & Red, Green, Blue \\
        HSI & Hue, Saturation, Intensity \\
        SVM & Support vector machine \\
       H\&E & Hematoxylin and Eosin \\
       EMC & Erasmus medical centre \\
       SUH & Stavanger University Hospital \\  
    \end{tabular}
    \label{tab:my_label}
\end{table} 

\section*{Declarations}
\subsection*{Availability of data and materials}
The code is available at \href{https://github.com/NeelKanwal/Equipping-Computational-Pathology-Systems-with-Artifact-Processing-Pipeline}{Github}. The training and development dataset (named HistoArtifacts) can be downloaded from \href{https://zenodo.org/records/10809442?token=eyJhbGciOiJIUzUxMiJ9.eyJpZCI6ImZjNjg0MDNjLTk3Y2UtNDYyNC1hZWVkLTFmNTc2MWE3Zjg2NCIsImRhdGEiOnt9LCJyYW5kb20iOiIwNjYwNjg3MTk1YzM2NzdiNmZjMTE1YzA2MWY1YWVhZiJ9.3QNjcwxXYAxQoLVuBShsUhDkGtiFOd7IPViM6O8hG3XL9-AzIhTm1_YqlYr-oX_fi5lW5-oZQ33FPTmKpfkZ9Q}{Zenodo}.

\subsection*{Ethics approval and consent to participate}
This study was performed in line with the principles of the Declaration of Helsinki. The patients/participants provided written informed consent to use their data for secondary purposes. The Erasmus MC Medical Research Committee granted approval from the Institutional Review Board under the reference MEC-2018-1097. The Stavanger University Hospital's data is approved by the Norwegian Regional Committee for Medical and Health Research Ethics under REC, 2010/1241. INCLIVA Biomedical Research Institute granted approval from the Research Ethics Committee (CEIm) of the Hospital Clínico Universitario of Valencia, Spain, under the reference CEIm-2020/114.

\subsection*{Competing interests}
This research work is objective and unbiased. The authors have no relevant financial or non-financial interests to disclose.

\subsection*{Authors' contributions}
\textbf{N.K:} Conceptualization, Methodology, Software, Validation, Investigation, Data curation, Writing - Original Draft, Writing - Review \& Editing, Visualization; \textbf{F.K:} Formal analysis; Investigation; Data Resources;  Writing - Review \& Editing; \textbf{U.K:} Formal analysis; Investigation; Data Resources;  Writing-Review \& Editing; \textbf{A.M:}  Formal analysis; Investigation; Data Resources;  Writing - Review \& Editing; \textbf{C.M:} Formal analysis; Data Resources; Writing - Review \& Editing; Funding acquisition; \textbf{E.J:} Formal analysis; Data Resources; Writing - Review \& Editing; Funding acquisition; \textbf{T.Z:}  Formal analysis; Data Resources; Writing - Review \& Editing; Funding acquisition; \textbf{C.R:} Writing - Review \& Editing; \textbf{K.E:} Conceptualization; Methodology; Investigation; Supervision; Writing - Review \& Editing; Project administration; Funding acquisition.

\subsection*{Consent for publication}
All authors consent to the open-access publication of this research work.

\subsection*{Acknowledgements and Funding}
The European Union's Horizon 2020 research and innovation program (CLARIFY) financially supports this research work under the Marie Skłodowska-Curie grant agreement 860627.

\bibliography{refs}


\begin{thebibliography}{88}
\ifx \bisbn   \undefined \def \bisbn  #1{ISBN #1}\fi
\ifx \binits  \undefined \def \binits#1{#1}\fi
\ifx \bauthor  \undefined \def \bauthor#1{#1}\fi
\ifx \batitle  \undefined \def \batitle#1{#1}\fi
\ifx \bjtitle  \undefined \def \bjtitle#1{#1}\fi
\ifx \bvolume  \undefined \def \bvolume#1{\textbf{#1}}\fi
\ifx \byear  \undefined \def \byear#1{#1}\fi
\ifx \bissue  \undefined \def \bissue#1{#1}\fi
\ifx \bfpage  \undefined \def \bfpage#1{#1}\fi
\ifx \blpage  \undefined \def \blpage #1{#1}\fi
\ifx \burl  \undefined \def \burl#1{\textsf{#1}}\fi
\ifx \doiurl  \undefined \def \doiurl#1{\url{https://doi.org/#1}}\fi
\ifx \betal  \undefined \def \betal{\textit{et al.}}\fi
\ifx \binstitute  \undefined \def \binstitute#1{#1}\fi
\ifx \binstitutionaled  \undefined \def \binstitutionaled#1{#1}\fi
\ifx \bctitle  \undefined \def \bctitle#1{#1}\fi
\ifx \beditor  \undefined \def \beditor#1{#1}\fi
\ifx \bpublisher  \undefined \def \bpublisher#1{#1}\fi
\ifx \bbtitle  \undefined \def \bbtitle#1{#1}\fi
\ifx \bedition  \undefined \def \bedition#1{#1}\fi
\ifx \bseriesno  \undefined \def \bseriesno#1{#1}\fi
\ifx \blocation  \undefined \def \blocation#1{#1}\fi
\ifx \bsertitle  \undefined \def \bsertitle#1{#1}\fi
\ifx \bsnm \undefined \def \bsnm#1{#1}\fi
\ifx \bsuffix \undefined \def \bsuffix#1{#1}\fi
\ifx \bparticle \undefined \def \bparticle#1{#1}\fi
\ifx \barticle \undefined \def \barticle#1{#1}\fi
\bibcommenthead
\ifx \bconfdate \undefined \def \bconfdate #1{#1}\fi
\ifx \botherref \undefined \def \botherref #1{#1}\fi
\ifx \url \undefined \def \url#1{\textsf{#1}}\fi
\ifx \bchapter \undefined \def \bchapter#1{#1}\fi
\ifx \bbook \undefined \def \bbook#1{#1}\fi
\ifx \bcomment \undefined \def \bcomment#1{#1}\fi
\ifx \oauthor \undefined \def \oauthor#1{#1}\fi
\ifx \citeauthoryear \undefined \def \citeauthoryear#1{#1}\fi
\ifx \endbibitem  \undefined \def \endbibitem {}\fi
\ifx \bconflocation  \undefined \def \bconflocation#1{#1}\fi
\ifx \arxivurl  \undefined \def \arxivurl#1{\textsf{#1}}\fi
\csname PreBibitemsHook\endcsname

\bibitem[\protect\citeauthoryear{{National Cancer Institute}}{2015}]{cancer-gov}
\begin{botherref}
\oauthor{\bsnm{{National Cancer Institute}}}:
Environmental Carcinogens and Cancer Risk.
\url{https://www.cancer.gov/about-cancer/causes-prevention/risk/substances/carcinogens}.
Accessed on August 31, 2023
(2015)
\end{botherref}
\endbibitem

\bibitem[\protect\citeauthoryear{{World Cancer Research Fund International}}{2023}]{wcrf_stats}
\begin{botherref}
\oauthor{\bsnm{{World Cancer Research Fund International}}}:
Differences in cancer incidence and mortality across the globe.
\url{https://www.wcrf.org/differences-in-cancer-incidence-and-mortality-across-the-globe/}.
Accessed on August 31, 2023
(2023)
\end{botherref}
\endbibitem

\bibitem[\protect\citeauthoryear{Pulumati et~al.}{2023}]{pulumati2023technological}
\begin{barticle}
\bauthor{\bsnm{Pulumati}, \binits{A.}},
\bauthor{\bsnm{Pulumati}, \binits{A.}},
\bauthor{\bsnm{Dwarakanath}, \binits{B.S.}},
\bauthor{\bsnm{Verma}, \binits{A.}},
\bauthor{\bsnm{Papineni}, \binits{R.V.}}:
\batitle{Technological advancements in cancer diagnostics: Improvements and limitations}.
\bjtitle{Cancer Reports}
\bvolume{6}(\bissue{2}),
\bfpage{1764}
(\byear{2023})
\end{barticle}
\endbibitem

\bibitem[\protect\citeauthoryear{Khened et~al.}{2021}]{khened2021generalized}
\begin{barticle}
\bauthor{\bsnm{Khened}, \binits{M.}},
\bauthor{\bsnm{Kori}, \binits{A.}},
\bauthor{\bsnm{Rajkumar}, \binits{H.}},
\bauthor{\bsnm{Krishnamurthi}, \binits{G.}},
\bauthor{\bsnm{Srinivasan}, \binits{B.}}:
\batitle{A generalized deep learning framework for whole-slide image segmentation and analysis}.
\bjtitle{Scientific reports}
\bvolume{11}(\bissue{1}),
\bfpage{11579}
(\byear{2021})
\end{barticle}
\endbibitem

\bibitem[\protect\citeauthoryear{Zhu et~al.}{2019}]{zhu2019breast}
\begin{barticle}
\bauthor{\bsnm{Zhu}, \binits{C.}},
\bauthor{\bsnm{Song}, \binits{F.}},
\bauthor{\bsnm{Wang}, \binits{Y.}},
\bauthor{\bsnm{Dong}, \binits{H.}},
\bauthor{\bsnm{Guo}, \binits{Y.}},
\bauthor{\bsnm{Liu}, \binits{J.}}:
\batitle{Breast cancer histopathology image classification through assembling multiple compact cnns}.
\bjtitle{BMC medical informatics and decision making}
\bvolume{19}(\bissue{1}),
\bfpage{1}--\blpage{17}
(\byear{2019})
\end{barticle}
\endbibitem

\bibitem[\protect\citeauthoryear{Kanwal et~al.}{2023}]{kanwal2023detection}
\begin{bchapter}
\bauthor{\bsnm{Kanwal}, \binits{N.}},
\bauthor{\bsnm{Amundsen}, \binits{R.}},
\bauthor{\bsnm{Hardardottir}, \binits{H.}},
\bauthor{\bsnm{Janssen}, \binits{E.A.}},
\bauthor{\bsnm{Engan}, \binits{K.}}:
\bctitle{Detection and localization of melanoma skin cancer in histopathological whole slide images}.
In: \bbtitle{2023 31st European Signal Processing Conference (EUSIPCO)},
pp. \bfpage{1128}--\blpage{1135}
(\byear{2023}).
\bcomment{IEEE}
\end{bchapter}
\endbibitem

\bibitem[\protect\citeauthoryear{Car et~al.}{2016}]{car2016clinician}
\begin{barticle}
\bauthor{\bsnm{Car}, \binits{L.T.}},
\bauthor{\bsnm{Papachristou}, \binits{N.}},
\bauthor{\bsnm{Bull}, \binits{A.}},
\bauthor{\bsnm{Majeed}, \binits{A.}},
\bauthor{\bsnm{Gallagher}, \binits{J.}},
\bauthor{\bsnm{El-Khatib}, \binits{M.}},
\bauthor{\bsnm{Aylin}, \binits{P.}},
\bauthor{\bsnm{Rudan}, \binits{I.}},
\bauthor{\bsnm{Atun}, \binits{R.}},
\bauthor{\bsnm{Car}, \binits{J.}}, \betal:
\batitle{Clinician-identified problems and solutions for delayed diagnosis in primary care: a prioritize study}.
\bjtitle{BMC family practice}
\bvolume{17},
\bfpage{1}--\blpage{9}
(\byear{2016})
\end{barticle}
\endbibitem

\bibitem[\protect\citeauthoryear{Pallua et~al.}{2020}]{pallua2020future}
\begin{barticle}
\bauthor{\bsnm{Pallua}, \binits{J.}},
\bauthor{\bsnm{Brunner}, \binits{A.}},
\bauthor{\bsnm{Zelger}, \binits{B.}},
\bauthor{\bsnm{Schirmer}, \binits{M.}},
\bauthor{\bsnm{Haybaeck}, \binits{J.}}:
\batitle{The future of pathology is digital}.
\bjtitle{Pathology-Research and Practice}
\bvolume{216}(\bissue{9}),
\bfpage{153040}
(\byear{2020})
\end{barticle}
\endbibitem

\bibitem[\protect\citeauthoryear{Inc}{}]{dimensionsai}
\begin{botherref}
\oauthor{\bsnm{Inc}, \binits{D.S..R.S.}}:
Digital Science and Research Solutions Inc.
\url{https://app.dimensions.ai/analytics/publication/overview/timeline?search_mode=content&or_facet_year=2018&or_facet_year=2019&or_facet_year=2020&or_facet_year=2021&or_facet_year=2022&or_facet_year=2023&search_text=Digital%20Pathology&search_type=kws&search_field=full_search}.
Query: "CPATH" OR "Computational Pathology" OR "Digital Pathology" (accessed: August 2023)
\end{botherref}
\endbibitem

\bibitem[\protect\citeauthoryear{Kanwal et~al.}{2022}]{kanwal2022devil}
\begin{botherref}
\oauthor{\bsnm{Kanwal}, \binits{N.}},
\oauthor{\bsnm{P{\'e}rez-Bueno}, \binits{F.}},
\oauthor{\bsnm{Schmidt}, \binits{A.}},
\oauthor{\bsnm{Molina}, \binits{R.}},
\oauthor{\bsnm{Engan}, \binits{K.}}:
The devil is in the details: Whole slide image acquisition and processing for artifacts detection, color variation, and data augmentation. a review.
IEEE Access
(2022)
\end{botherref}
\endbibitem

\bibitem[\protect\citeauthoryear{Campanella et~al.}{2018}]{CAMPANELLA2018142}
\begin{barticle}
\bauthor{\bsnm{Campanella}, \binits{G.}},
\bauthor{\bsnm{Rajanna}, \binits{A.R.}},
\bauthor{\bsnm{Corsale}, \binits{L.}},
\bauthor{\bsnm{Schüffler}, \binits{P.J.}},
\bauthor{\bsnm{Yagi}, \binits{Y.}},
\bauthor{\bsnm{Fuchs}, \binits{T.J.}}:
\batitle{Towards machine learned quality control: A benchmark for sharpness quantification in digital pathology}.
\bjtitle{Computerized Medical Imaging and Graphics}
\bvolume{65},
\bfpage{142}--\blpage{151}
(\byear{2018})
\end{barticle}
\endbibitem

\bibitem[\protect\citeauthoryear{Hosseini et~al.}{2024}]{hosseini2024computational}
\begin{botherref}
\oauthor{\bsnm{Hosseini}, \binits{M.S.}},
\oauthor{\bsnm{Bejnordi}, \binits{B.E.}},
\oauthor{\bsnm{Trinh}, \binits{V.Q.-H.}},
\oauthor{\bsnm{Chan}, \binits{L.}},
\oauthor{\bsnm{Hasan}, \binits{D.}},
\oauthor{\bsnm{Li}, \binits{X.}},
\oauthor{\bsnm{Yang}, \binits{S.}},
\oauthor{\bsnm{Kim}, \binits{T.}},
\oauthor{\bsnm{Zhang}, \binits{H.}},
\oauthor{\bsnm{Wu}, \binits{T.}}, et al.:
Computational pathology: a survey review and the way forward.
Journal of Pathology Informatics,
100357
(2024)
\end{botherref}
\endbibitem

\bibitem[\protect\citeauthoryear{Louis et~al.}{2014}]{louis2014computational}
\begin{barticle}
\bauthor{\bsnm{Louis}, \binits{D.N.}},
\bauthor{\bsnm{Gerber}, \binits{G.K.}},
\bauthor{\bsnm{Baron}, \binits{J.M.}},
\bauthor{\bsnm{Bry}, \binits{L.}},
\bauthor{\bsnm{Dighe}, \binits{A.S.}},
\bauthor{\bsnm{Getz}, \binits{G.}},
\bauthor{\bsnm{Higgins}, \binits{J.M.}},
\bauthor{\bsnm{Kuo}, \binits{F.C.}},
\bauthor{\bsnm{Lane}, \binits{W.J.}},
\bauthor{\bsnm{Michaelson}, \binits{J.S.}}, \betal:
\batitle{Computational pathology: an emerging definition}.
\bjtitle{Archives of pathology \& laboratory medicine}
\bvolume{138}(\bissue{9}),
\bfpage{1133}--\blpage{1138}
(\byear{2014})
\end{barticle}
\endbibitem

\bibitem[\protect\citeauthoryear{Taqi et~al.}{2018}]{taqi2018review}
\begin{barticle}
\bauthor{\bsnm{Taqi}, \binits{S.A.}},
\bauthor{\bsnm{Sami}, \binits{S.A.}},
\bauthor{\bsnm{Sami}, \binits{L.B.}},
\bauthor{\bsnm{Zaki}, \binits{S.A.}}:
\batitle{A review of artifacts in histopathology}.
\bjtitle{Journal of oral and maxillofacial pathology: JOMFP}
\bvolume{22}(\bissue{2}),
\bfpage{279}
(\byear{2018})
\end{barticle}
\endbibitem

\bibitem[\protect\citeauthoryear{Bindhu et~al.}{2013}]{bindhu2013facts}
\begin{barticle}
\bauthor{\bsnm{Bindhu}, \binits{P.}},
\bauthor{\bsnm{Krishnapillai}, \binits{R.}},
\bauthor{\bsnm{Thomas}, \binits{P.}},
\bauthor{\bsnm{Jayanthi}, \binits{P.}}:
\batitle{Facts in artifacts}.
\bjtitle{Journal of oral and maxillofacial pathology: JOMFP}
\bvolume{17}(\bissue{3}),
\bfpage{397}
(\byear{2013})
\end{barticle}
\endbibitem

\bibitem[\protect\citeauthoryear{Kanwal et~al.}{2023}]{kanwal2023vision}
\begin{bchapter}
\bauthor{\bsnm{Kanwal}, \binits{N.}},
\bauthor{\bsnm{Eftest{\o}l}, \binits{T.}},
\bauthor{\bsnm{Khoraminia}, \binits{F.}},
\bauthor{\bsnm{Zuiverloon}, \binits{T.C.}},
\bauthor{\bsnm{Engan}, \binits{K.}}:
\bctitle{Vision transformers for small histological datasets learned through knowledge distillation}.
In: \bbtitle{Pacific-Asia Conference on Knowledge Discovery and Data Mining},
pp. \bfpage{167}--\blpage{179}
(\byear{2023}).
\bcomment{Springer}
\end{bchapter}
\endbibitem

\bibitem[\protect\citeauthoryear{Wright et~al.}{2020}]{wright2020effect}
\begin{barticle}
\bauthor{\bsnm{Wright}, \binits{A.I.}},
\bauthor{\bsnm{Dunn}, \binits{C.M.}},
\bauthor{\bsnm{Hale}, \binits{M.}},
\bauthor{\bsnm{Hutchins}, \binits{G.G.}},
\bauthor{\bsnm{Treanor}, \binits{D.E.}}:
\batitle{The effect of quality control on accuracy of digital pathology image analysis}.
\bjtitle{IEEE Journal of Biomedical and Health Informatics}
\bvolume{25}(\bissue{2}),
\bfpage{307}--\blpage{314}
(\byear{2020})
\end{barticle}
\endbibitem

\bibitem[\protect\citeauthoryear{Tabatabaei et~al.}{2022}]{tabatabaei2022residual}
\begin{bchapter}
\bauthor{\bsnm{Tabatabaei}, \binits{Z.}},
\bauthor{\bsnm{Colomer}, \binits{A.}},
\bauthor{\bsnm{Engan}, \binits{K.}},
\bauthor{\bsnm{Oliver}, \binits{J.}},
\bauthor{\bsnm{Naranjo}, \binits{V.}}:
\bctitle{Residual block convolutional auto encoder in content-based medical image retrieval}.
In: \bbtitle{2022 IEEE 14th Image, Video, and Multidimensional Signal Processing Workshop (IVMSP)},
pp. \bfpage{1}--\blpage{5}
(\byear{2022}).
\bcomment{IEEE}
\end{bchapter}
\endbibitem

\bibitem[\protect\citeauthoryear{Chen et~al.}{2022}]{chen2022classification}
\begin{barticle}
\bauthor{\bsnm{Chen}, \binits{C.}},
\bauthor{\bsnm{Chen}, \binits{C.}},
\bauthor{\bsnm{Ma}, \binits{M.}},
\bauthor{\bsnm{Ma}, \binits{X.}},
\bauthor{\bsnm{Lv}, \binits{X.}},
\bauthor{\bsnm{Dong}, \binits{X.}},
\bauthor{\bsnm{Yan}, \binits{Z.}},
\bauthor{\bsnm{Zhu}, \binits{M.}},
\bauthor{\bsnm{Chen}, \binits{J.}}:
\batitle{Classification of multi-differentiated liver cancer pathological images based on deep learning attention mechanism}.
\bjtitle{BMC Medical Informatics and Decision Making}
\bvolume{22}(\bissue{1}),
\bfpage{1}--\blpage{13}
(\byear{2022})
\end{barticle}
\endbibitem

\bibitem[\protect\citeauthoryear{Fuster et~al.}{2022}]{9816352}
\begin{bchapter}
\bauthor{\bsnm{Fuster}, \binits{S.}},
\bauthor{\bsnm{Khoraminia}, \binits{F.}},
\bauthor{\bsnm{Kiraz}, \binits{U.}},
\bauthor{\bsnm{Kanwal}, \binits{N.}},
\bauthor{\bsnm{Kvikstad}, \binits{V.}},
\bauthor{\bsnm{Eftestøl}, \binits{T.}},
\bauthor{\bsnm{Zuiverloon}, \binits{T.C.M.}},
\bauthor{\bsnm{Janssen}, \binits{E.A.M.}},
\bauthor{\bsnm{Engan}, \binits{K.}}:
\bctitle{Invasive cancerous area detection in non-muscle invasive bladder cancer whole slide images}.
In: \bbtitle{2022 IEEE 14th Image, Video, and Multidimensional Signal Processing Workshop (IVMSP)},
pp. \bfpage{1}--\blpage{5}
(\byear{2022})
\end{bchapter}
\endbibitem

\bibitem[\protect\citeauthoryear{Litjens et~al.}{2017}]{litjens2017survey}
\begin{barticle}
\bauthor{\bsnm{Litjens}, \binits{G.}},
\bauthor{\bsnm{Kooi}, \binits{T.}},
\bauthor{\bsnm{Bejnordi}, \binits{B.E.}},
\bauthor{\bsnm{Setio}, \binits{A.A.A.}},
\bauthor{\bsnm{Ciompi}, \binits{F.}},
\bauthor{\bsnm{Ghafoorian}, \binits{M.}},
\bauthor{\bsnm{Van Der~Laak}, \binits{J.A.}},
\bauthor{\bsnm{Van~Ginneken}, \binits{B.}},
\bauthor{\bsnm{S{\'a}nchez}, \binits{C.I.}}:
\batitle{A survey on deep learning in medical image analysis}.
\bjtitle{Medical image analysis}
\bvolume{42},
\bfpage{60}--\blpage{88}
(\byear{2017})
\end{barticle}
\endbibitem

\bibitem[\protect\citeauthoryear{Chen et~al.}{2021}]{chen2021review}
\begin{barticle}
\bauthor{\bsnm{Chen}, \binits{L.}},
\bauthor{\bsnm{Li}, \binits{S.}},
\bauthor{\bsnm{Bai}, \binits{Q.}},
\bauthor{\bsnm{Yang}, \binits{J.}},
\bauthor{\bsnm{Jiang}, \binits{S.}},
\bauthor{\bsnm{Miao}, \binits{Y.}}:
\batitle{Review of image classification algorithms based on convolutional neural networks}.
\bjtitle{Remote Sensing}
\bvolume{13}(\bissue{22}),
\bfpage{4712}
(\byear{2021})
\end{barticle}
\endbibitem

\bibitem[\protect\citeauthoryear{Lu et~al.}{2022}]{lu2022bridging}
\begin{barticle}
\bauthor{\bsnm{Lu}, \binits{Z.}},
\bauthor{\bsnm{Xie}, \binits{H.}},
\bauthor{\bsnm{Liu}, \binits{C.}},
\bauthor{\bsnm{Zhang}, \binits{Y.}}:
\batitle{Bridging the gap between vision transformers and convolutional neural networks on small datasets}.
\bjtitle{Advances in Neural Information Processing Systems}
\bvolume{35},
\bfpage{14663}--\blpage{14677}
(\byear{2022})
\end{barticle}
\endbibitem

\bibitem[\protect\citeauthoryear{Zhu et~al.}{2023}]{zhu2023understanding}
\begin{botherref}
\oauthor{\bsnm{Zhu}, \binits{H.}},
\oauthor{\bsnm{Chen}, \binits{B.}},
\oauthor{\bsnm{Yang}, \binits{C.}}:
Understanding why vit trains badly on small datasets: An intuitive perspective.
arXiv preprint arXiv:2302.03751
(2023)
\end{botherref}
\endbibitem

\bibitem[\protect\citeauthoryear{Atabansi et~al.}{2023}]{atabansi2023survey}
\begin{barticle}
\bauthor{\bsnm{Atabansi}, \binits{C.C.}},
\bauthor{\bsnm{Nie}, \binits{J.}},
\bauthor{\bsnm{Liu}, \binits{H.}},
\bauthor{\bsnm{Song}, \binits{Q.}},
\bauthor{\bsnm{Yan}, \binits{L.}},
\bauthor{\bsnm{Zhou}, \binits{X.}}:
\batitle{A survey of transformer applications for histopathological image analysis: New developments and future directions}.
\bjtitle{BioMedical Engineering OnLine}
\bvolume{22}(\bissue{1}),
\bfpage{96}
(\byear{2023})
\end{barticle}
\endbibitem

\bibitem[\protect\citeauthoryear{Naseer et~al.}{2021}]{naseer2021intriguing}
\begin{barticle}
\bauthor{\bsnm{Naseer}, \binits{M.M.}},
\bauthor{\bsnm{Ranasinghe}, \binits{K.}},
\bauthor{\bsnm{Khan}, \binits{S.H.}},
\bauthor{\bsnm{Hayat}, \binits{M.}},
\bauthor{\bsnm{Shahbaz~Khan}, \binits{F.}},
\bauthor{\bsnm{Yang}, \binits{M.-H.}}:
\batitle{Intriguing properties of vision transformers}.
\bjtitle{Advances in Neural Information Processing Systems}
\bvolume{34},
\bfpage{23296}--\blpage{23308}
(\byear{2021})
\end{barticle}
\endbibitem

\bibitem[\protect\citeauthoryear{Bhojanapalli et~al.}{2021}]{bhojanapalli2021}
\begin{bchapter}
\bauthor{\bsnm{Bhojanapalli}, \binits{S.}},
\bauthor{\bsnm{Chakrabarti}, \binits{A.}},
\bauthor{\bsnm{Glasner}, \binits{D.}},
\bauthor{\bsnm{Li}, \binits{D.}},
\bauthor{\bsnm{Unterthiner}, \binits{T.}},
\bauthor{\bsnm{Veit}, \binits{A.}}:
\bctitle{Understanding robustness of transformers for image classification}.
In: \bbtitle{Proceedings of the IEEE/CVF International Conference on Computer Vision},
pp. \bfpage{10231}--\blpage{10241}
(\byear{2021})
\end{bchapter}
\endbibitem

\bibitem[\protect\citeauthoryear{Hsu et~al.}{2022}]{hsu2022automatic}
\begin{barticle}
\bauthor{\bsnm{Hsu}, \binits{S.-T.}},
\bauthor{\bsnm{Su}, \binits{Y.-J.}},
\bauthor{\bsnm{Hung}, \binits{C.-H.}},
\bauthor{\bsnm{Chen}, \binits{M.-J.}},
\bauthor{\bsnm{Lu}, \binits{C.-H.}},
\bauthor{\bsnm{Kuo}, \binits{C.-E.}}:
\batitle{Automatic ovarian tumors recognition system based on ensemble convolutional neural network with ultrasound imaging}.
\bjtitle{BMC Medical Informatics and Decision Making}
\bvolume{22}(\bissue{1}),
\bfpage{298}
(\byear{2022})
\end{barticle}
\endbibitem

\bibitem[\protect\citeauthoryear{Meng et~al.}{2021}]{meng:2021}
\begin{barticle}
\bauthor{\bsnm{Meng}, \binits{Z.}},
\bauthor{\bsnm{Zhao}, \binits{Z.}},
\bauthor{\bsnm{Li}, \binits{B.}},
\bauthor{\bsnm{Su}, \binits{F.}},
\bauthor{\bsnm{Guo}, \binits{L.}}:
\batitle{A cervical histopathology dataset for computer aided diagnosis of precancerous lesions}.
\bjtitle{IEEE Transactions on Medical Imaging}
\bvolume{40}(\bissue{6}),
\bfpage{1531}--\blpage{1541}
(\byear{2021})
\end{barticle}
\endbibitem

\bibitem[\protect\citeauthoryear{Abe et~al.}{2022}]{abe2022deep}
\begin{barticle}
\bauthor{\bsnm{Abe}, \binits{T.}},
\bauthor{\bsnm{Buchanan}, \binits{E.K.}},
\bauthor{\bsnm{Pleiss}, \binits{G.}},
\bauthor{\bsnm{Zemel}, \binits{R.}},
\bauthor{\bsnm{Cunningham}, \binits{J.P.}}:
\batitle{Deep ensembles work, but are they necessary?}
\bjtitle{Advances in Neural Information Processing Systems}
\bvolume{35},
\bfpage{33646}--\blpage{33660}
(\byear{2022})
\end{barticle}
\endbibitem

\bibitem[\protect\citeauthoryear{Mohammed and Kora}{2023}]{mohammed2023comprehensive}
\begin{botherref}
\oauthor{\bsnm{Mohammed}, \binits{A.}},
\oauthor{\bsnm{Kora}, \binits{R.}}:
A comprehensive review on ensemble deep learning: Opportunities and challenges.
Journal of King Saud University-Computer and Information Sciences
(2023)
\end{botherref}
\endbibitem

\bibitem[\protect\citeauthoryear{Howard et~al.}{2019}]{mobilenet}
\begin{bchapter}
\bauthor{\bsnm{Howard}, \binits{A.}},
\bauthor{\bsnm{Sandler}, \binits{M.}},
\bauthor{\bsnm{Chu}, \binits{G.}},
\bauthor{\bsnm{Chen}, \binits{L.-C.}},
\bauthor{\bsnm{Chen}, \binits{B.}},
\bauthor{\bsnm{Tan}, \binits{M.}},
\bauthor{\bsnm{Wang}, \binits{W.}},
\bauthor{\bsnm{Zhu}, \binits{Y.}},
\bauthor{\bsnm{Pang}, \binits{R.}},
\bauthor{\bsnm{Vasudevan}, \binits{V.}}, \betal:
\bctitle{Searching for mobilenetv3}.
In: \bbtitle{Proceedings of the IEEE/CVF International Conference on Computer Vision},
pp. \bfpage{1314}--\blpage{1324}
(\byear{2019})
\end{bchapter}
\endbibitem

\bibitem[\protect\citeauthoryear{Touvron et~al.}{2021}]{Deit}
\begin{bchapter}
\bauthor{\bsnm{Touvron}, \binits{H.}},
\bauthor{\bsnm{Cord}, \binits{M.}},
\bauthor{\bsnm{Douze}, \binits{M.}},
\bauthor{\bsnm{Massa}, \binits{F.}},
\bauthor{\bsnm{Sablayrolles}, \binits{A.}},
\bauthor{\bsnm{J{\'e}gou}, \binits{H.}}:
\bctitle{Training data-efficient image transformers \& distillation through attention}.
In: \bbtitle{International Conference on Machine Learning},
pp. \bfpage{10347}--\blpage{10357}
(\byear{2021}).
\bcomment{PMLR}
\end{bchapter}
\endbibitem

\bibitem[\protect\citeauthoryear{Morales et~al.}{2021}]{morales2021artificial}
\begin{barticle}
\bauthor{\bsnm{Morales}, \binits{S.}},
\bauthor{\bsnm{Engan}, \binits{K.}},
\bauthor{\bsnm{Naranjo}, \binits{V.}}:
\batitle{Artificial intelligence in computational pathology--challenges and future directions}.
\bjtitle{Digital Signal Processing}
\bvolume{119},
\bfpage{103196}
(\byear{2021})
\end{barticle}
\endbibitem

\bibitem[\protect\citeauthoryear{Bulten et~al.}{2022}]{bulten2022artificial}
\begin{barticle}
\bauthor{\bsnm{Bulten}, \binits{W.}},
\bauthor{\bsnm{Kartasalo}, \binits{K.}},
\bauthor{\bsnm{Chen}, \binits{P.-H.C.}},
\bauthor{\bsnm{Str{\"o}m}, \binits{P.}},
\bauthor{\bsnm{Pinckaers}, \binits{H.}},
\bauthor{\bsnm{Nagpal}, \binits{K.}},
\bauthor{\bsnm{Cai}, \binits{Y.}},
\bauthor{\bsnm{Steiner}, \binits{D.F.}},
\bauthor{\bsnm{Boven}, \binits{H.}},
\bauthor{\bsnm{Vink}, \binits{R.}}, \betal:
\batitle{Artificial intelligence for diagnosis and gleason grading of prostate cancer: the panda challenge}.
\bjtitle{Nature medicine}
\bvolume{28}(\bissue{1}),
\bfpage{154}--\blpage{163}
(\byear{2022})
\end{barticle}
\endbibitem

\bibitem[\protect\citeauthoryear{Khoraminia et~al.}{2023}]{khoraminia2023artificial}
\begin{barticle}
\bauthor{\bsnm{Khoraminia}, \binits{F.}},
\bauthor{\bsnm{Fuster}, \binits{S.}},
\bauthor{\bsnm{Kanwal}, \binits{N.}},
\bauthor{\bsnm{Olislagers}, \binits{M.}},
\bauthor{\bsnm{Engan}, \binits{K.}},
\bauthor{\bsnm{Leenders}, \binits{G.J.}},
\bauthor{\bsnm{Stubbs}, \binits{A.P.}},
\bauthor{\bsnm{Akram}, \binits{F.}},
\bauthor{\bsnm{Zuiverloon}, \binits{T.C.}}:
\batitle{Artificial intelligence in digital pathology for bladder cancer: Hype or hope? a systematic review}.
\bjtitle{Cancers}
\bvolume{15}(\bissue{18}),
\bfpage{4518}
(\byear{2023})
\end{barticle}
\endbibitem

\bibitem[\protect\citeauthoryear{Gay et~al.}{2019}]{gay2019texture}
\begin{bchapter}
\bauthor{\bsnm{Gay}, \binits{J.}},
\bauthor{\bsnm{Harlin}, \binits{H.}},
\bauthor{\bsnm{Wetzer}, \binits{E.}},
\bauthor{\bsnm{Lindblad}, \binits{J.}},
\bauthor{\bsnm{Sladoje}, \binits{N.}}:
\bctitle{Texture-based oral cancer detection: A performance analysis of deep learning approaches.}
In: \bbtitle{3rd NEUBIAS Conference}
(\byear{2019})
\end{bchapter}
\endbibitem

\bibitem[\protect\citeauthoryear{Gandomkar et~al.}{2018}]{GANDOMKAR201814}
\begin{barticle}
\bauthor{\bsnm{Gandomkar}, \binits{Z.}},
\bauthor{\bsnm{Brennan}, \binits{P.C.}},
\bauthor{\bsnm{Mello-Thoms}, \binits{C.}}:
\batitle{Mudern: Multi-category classification of breast histopathological image using deep residual networks}.
\bjtitle{Artificial Intelligence in Medicine}
\bvolume{88},
\bfpage{14}--\blpage{24}
(\byear{2018})
\doiurl{10.1016/j.artmed.2018.04.005}
\end{barticle}
\endbibitem

\bibitem[\protect\citeauthoryear{Wessels et~al.}{2023}]{wessels2023self}
\begin{barticle}
\bauthor{\bsnm{Wessels}, \binits{F.}},
\bauthor{\bsnm{Schmitt}, \binits{M.}},
\bauthor{\bsnm{Krieghoff-Henning}, \binits{E.}},
\bauthor{\bsnm{Nientiedt}, \binits{M.}},
\bauthor{\bsnm{Waldbillig}, \binits{F.}},
\bauthor{\bsnm{Neuberger}, \binits{M.}},
\bauthor{\bsnm{Kriegmair}, \binits{M.C.}},
\bauthor{\bsnm{Kowalewski}, \binits{K.-F.}},
\bauthor{\bsnm{Worst}, \binits{T.S.}},
\bauthor{\bsnm{Steeg}, \binits{M.}}, \betal:
\batitle{A self-supervised vision transformer to predict survival from histopathology in renal cell carcinoma}.
\bjtitle{World Journal of Urology}
\bvolume{41}(\bissue{8}),
\bfpage{2233}--\blpage{2241}
(\byear{2023})
\end{barticle}
\endbibitem

\bibitem[\protect\citeauthoryear{Stegm{\"u}ller et~al.}{2023}]{stegmuller2023scorenet}
\begin{bchapter}
\bauthor{\bsnm{Stegm{\"u}ller}, \binits{T.}},
\bauthor{\bsnm{Bozorgtabar}, \binits{B.}},
\bauthor{\bsnm{Spahr}, \binits{A.}},
\bauthor{\bsnm{Thiran}, \binits{J.-P.}}:
\bctitle{Scorenet: Learning non-uniform attention and augmentation for transformer-based histopathological image classification}.
In: \bbtitle{Proceedings of the IEEE/CVF Winter Conference on Applications of Computer Vision},
pp. \bfpage{6170}--\blpage{6179}
(\byear{2023})
\end{bchapter}
\endbibitem

\bibitem[\protect\citeauthoryear{Perincheri et~al.}{2021}]{perincheri2021independent}
\begin{barticle}
\bauthor{\bsnm{Perincheri}, \binits{S.}},
\bauthor{\bsnm{Levi}, \binits{A.W.}},
\bauthor{\bsnm{Celli}, \binits{R.}},
\bauthor{\bsnm{Gershkovich}, \binits{P.}},
\bauthor{\bsnm{Rimm}, \binits{D.}},
\bauthor{\bsnm{Morrow}, \binits{J.S.}},
\bauthor{\bsnm{Rothrock}, \binits{B.}},
\bauthor{\bsnm{Raciti}, \binits{P.}},
\bauthor{\bsnm{Klimstra}, \binits{D.}},
\bauthor{\bsnm{Sinard}, \binits{J.}}:
\batitle{An independent assessment of an artificial intelligence system for prostate cancer detection shows strong diagnostic accuracy}.
\bjtitle{Modern Pathology}
\bvolume{34}(\bissue{8}),
\bfpage{1588}--\blpage{1595}
(\byear{2021})
\end{barticle}
\endbibitem

\bibitem[\protect\citeauthoryear{Huang et~al.}{2017}]{huang2017densely}
\begin{bchapter}
\bauthor{\bsnm{Huang}, \binits{G.}},
\bauthor{\bsnm{Liu}, \binits{Z.}},
\bauthor{\bsnm{Van Der~Maaten}, \binits{L.}},
\bauthor{\bsnm{Weinberger}, \binits{K.Q.}}:
\bctitle{Densely connected convolutional networks}.
In: \bbtitle{Proceedings of the IEEE Conference on Computer Vision and Pattern Recognition},
pp. \bfpage{4700}--\blpage{4708}
(\byear{2017})
\end{bchapter}
\endbibitem

\bibitem[\protect\citeauthoryear{He et~al.}{2016}]{resnet}
\begin{bchapter}
\bauthor{\bsnm{He}, \binits{K.}},
\bauthor{\bsnm{Zhang}, \binits{X.}},
\bauthor{\bsnm{Ren}, \binits{S.}},
\bauthor{\bsnm{Sun}, \binits{J.}}:
\bctitle{Deep residual learning for image recognition}.
In: \bbtitle{Proceedings of the IEEE Conference on Computer Vision and Pattern Recognition},
pp. \bfpage{770}--\blpage{778}
(\byear{2016})
\end{bchapter}
\endbibitem

\bibitem[\protect\citeauthoryear{Szegedy et~al.}{2015}]{googlenet}
\begin{bchapter}
\bauthor{\bsnm{Szegedy}, \binits{C.}},
\bauthor{\bsnm{Liu}, \binits{W.}},
\bauthor{\bsnm{Jia}, \binits{Y.}},
\bauthor{\bsnm{Sermanet}, \binits{P.}},
\bauthor{\bsnm{Reed}, \binits{S.}},
\bauthor{\bsnm{Anguelov}, \binits{D.}},
\bauthor{\bsnm{Erhan}, \binits{D.}},
\bauthor{\bsnm{Vanhoucke}, \binits{V.}},
\bauthor{\bsnm{Rabinovich}, \binits{A.}}:
\bctitle{Going deeper with convolutions}.
In: \bbtitle{2015 IEEE Conference on Computer Vision and Pattern Recognition (CVPR)},
pp. \bfpage{1}--\blpage{9}
(\byear{2015})
\end{bchapter}
\endbibitem

\bibitem[\protect\citeauthoryear{Caron et~al.}{2021}]{caron2021emerging}
\begin{bchapter}
\bauthor{\bsnm{Caron}, \binits{M.}},
\bauthor{\bsnm{Touvron}, \binits{H.}},
\bauthor{\bsnm{Misra}, \binits{I.}},
\bauthor{\bsnm{J{\'e}gou}, \binits{H.}},
\bauthor{\bsnm{Mairal}, \binits{J.}},
\bauthor{\bsnm{Bojanowski}, \binits{P.}},
\bauthor{\bsnm{Joulin}, \binits{A.}}:
\bctitle{Emerging properties in self-supervised vision transformers}.
In: \bbtitle{Proceedings of the IEEE/CVF International Conference on Computer Vision},
pp. \bfpage{9650}--\blpage{9660}
(\byear{2021})
\end{bchapter}
\endbibitem

\bibitem[\protect\citeauthoryear{Zidan et~al.}{2023}]{zidan2023swincup}
\begin{barticle}
\bauthor{\bsnm{Zidan}, \binits{U.}},
\bauthor{\bsnm{Gaber}, \binits{M.M.}},
\bauthor{\bsnm{Abdelsamea}, \binits{M.M.}}:
\batitle{Swincup: Cascaded swin transformer for histopathological structures segmentation in colorectal cancer}.
\bjtitle{Expert Systems with Applications}
\bvolume{216},
\bfpage{119452}
(\byear{2023})
\end{barticle}
\endbibitem

\bibitem[\protect\citeauthoryear{Srinidhi et~al.}{2020}]{srinidhi2020deep}
\begin{botherref}
\oauthor{\bsnm{Srinidhi}, \binits{C.L.}},
\oauthor{\bsnm{Ciga}, \binits{O.}},
\oauthor{\bsnm{Martel}, \binits{A.L.}}:
Deep neural network models for computational histopathology: A survey.
Medical Image Analysis,
101813
(2020)
\end{botherref}
\endbibitem

\bibitem[\protect\citeauthoryear{Riasatian et~al.}{2021}]{RIASATIAN2021102032}
\begin{barticle}
\bauthor{\bsnm{Riasatian}, \binits{A.}},
\bauthor{\bsnm{Babaie}, \binits{M.}},
\bauthor{\bsnm{Maleki}, \binits{D.}},
\bauthor{\bsnm{Kalra}, \binits{S.}},
\bauthor{\bsnm{Valipour}, \binits{M.}},
\bauthor{\bsnm{Hemati}, \binits{S.}},
\bauthor{\bsnm{Zaveri}, \binits{M.}},
\bauthor{\bsnm{Safarpoor}, \binits{A.}},
\bauthor{\bsnm{Shafiei}, \binits{S.}},
\bauthor{\bsnm{Afshari}, \binits{M.}},
\bauthor{\bsnm{Rasoolijaberi}, \binits{M.}},
\bauthor{\bsnm{Sikaroudi}, \binits{M.}},
\bauthor{\bsnm{Adnan}, \binits{M.}},
\bauthor{\bsnm{Shah}, \binits{S.}},
\bauthor{\bsnm{Choi}, \binits{C.}},
\bauthor{\bsnm{Damaskinos}, \binits{S.}},
\bauthor{\bsnm{Campbell}, \binits{C.J.}},
\bauthor{\bsnm{Diamandis}, \binits{P.}},
\bauthor{\bsnm{Pantanowitz}, \binits{L.}},
\bauthor{\bsnm{Kashani}, \binits{H.}},
\bauthor{\bsnm{Ghodsi}, \binits{A.}},
\bauthor{\bsnm{Tizhoosh}, \binits{H.R.}}:
\batitle{Fine-tuning and training of densenet for histopathology image representation using tcga diagnostic slides}.
\bjtitle{Medical Image Analysis}
\bvolume{70},
\bfpage{102032}
(\byear{2021})
\end{barticle}
\endbibitem

\bibitem[\protect\citeauthoryear{Talo}{2019}]{TALO2019101743}
\begin{barticle}
\bauthor{\bsnm{Talo}, \binits{M.}}:
\batitle{Automated classification of histopathology images using transfer learning}.
\bjtitle{Artificial Intelligence in Medicine}
\bvolume{101},
\bfpage{101743}
(\byear{2019})
\end{barticle}
\endbibitem

\bibitem[\protect\citeauthoryear{Wang et~al.}{2020}]{wang:2020}
\begin{barticle}
\bauthor{\bsnm{Wang}, \binits{Y.}},
\bauthor{\bsnm{Peng}, \binits{T.}},
\bauthor{\bsnm{Duan}, \binits{J.}},
\bauthor{\bsnm{Zhu}, \binits{C.}},
\bauthor{\bsnm{Liu}, \binits{J.}},
\bauthor{\bsnm{Ye}, \binits{J.}},
\bauthor{\bsnm{Jin}, \binits{M.}}:
\batitle{Pathological image classification based on hard example guided cnn}.
\bjtitle{IEEE Access}
\bvolume{8},
\bfpage{114249}--\blpage{114258}
(\byear{2020})
\end{barticle}
\endbibitem

\bibitem[\protect\citeauthoryear{Wang et~al.}{2022}]{WANG2022103451}
\begin{barticle}
\bauthor{\bsnm{Wang}, \binits{C.}},
\bauthor{\bsnm{Gong}, \binits{W.}},
\bauthor{\bsnm{Cheng}, \binits{J.}},
\bauthor{\bsnm{Qian}, \binits{Y.}}:
\batitle{Dblcnn: Dependency-based lightweight convolutional neural network for multi-classification of breast histopathology images}.
\bjtitle{Biomedical Signal Processing and Control}
\bvolume{73},
\bfpage{103451}
(\byear{2022})
\end{barticle}
\endbibitem

\bibitem[\protect\citeauthoryear{Gao et~al.}{2021}]{gao2021instance}
\begin{bchapter}
\bauthor{\bsnm{Gao}, \binits{Z.}},
\bauthor{\bsnm{Hong}, \binits{B.}},
\bauthor{\bsnm{Zhang}, \binits{X.}},
\bauthor{\bsnm{Li}, \binits{Y.}},
\bauthor{\bsnm{Jia}, \binits{C.}},
\bauthor{\bsnm{Wu}, \binits{J.}},
\bauthor{\bsnm{Wang}, \binits{C.}},
\bauthor{\bsnm{Meng}, \binits{D.}},
\bauthor{\bsnm{Li}, \binits{C.}}:
\bctitle{Instance-based vision transformer for subtyping of papillary renal cell carcinoma in histopathological image}.
In: \bbtitle{Medical Image Computing and Computer Assisted Intervention--MICCAI 2021: 24th International Conference, Strasbourg, France, September 27--October 1, 2021, Proceedings, Part VIII 24},
pp. \bfpage{299}--\blpage{308}
(\byear{2021}).
\bcomment{Springer}
\end{bchapter}
\endbibitem

\bibitem[\protect\citeauthoryear{Sch{\"o}mig-Markiefka et~al.}{2021}]{schomig2021quality}
\begin{barticle}
\bauthor{\bsnm{Sch{\"o}mig-Markiefka}, \binits{B.}},
\bauthor{\bsnm{Pryalukhin}, \binits{A.}},
\bauthor{\bsnm{Hulla}, \binits{W.}},
\bauthor{\bsnm{Bychkov}, \binits{A.}},
\bauthor{\bsnm{Fukuoka}, \binits{J.}},
\bauthor{\bsnm{Madabhushi}, \binits{A.}},
\bauthor{\bsnm{Achter}, \binits{V.}},
\bauthor{\bsnm{Nieroda}, \binits{L.}},
\bauthor{\bsnm{B{\"u}ttner}, \binits{R.}},
\bauthor{\bsnm{Quaas}, \binits{A.}}, \betal:
\batitle{Quality control stress test for deep learning-based diagnostic model in digital pathology}.
\bjtitle{Modern Pathology}
\bvolume{34}(\bissue{12}),
\bfpage{2098}--\blpage{2108}
(\byear{2021})
\end{barticle}
\endbibitem

\bibitem[\protect\citeauthoryear{Linmans et~al.}{2024}]{linmans2024diffusion}
\begin{barticle}
\bauthor{\bsnm{Linmans}, \binits{J.}},
\bauthor{\bsnm{Raya}, \binits{G.}},
\bauthor{\bsnm{Laak}, \binits{J.}},
\bauthor{\bsnm{Litjens}, \binits{G.}}:
\batitle{Diffusion models for out-of-distribution detection in digital pathology}.
\bjtitle{Medical Image Analysis}
\bvolume{93},
\bfpage{103088}
(\byear{2024})
\end{barticle}
\endbibitem

\bibitem[\protect\citeauthoryear{Ghaffari~Laleh et~al.}{2022}]{ghaffari2022adversarial}
\begin{barticle}
\bauthor{\bsnm{Ghaffari~Laleh}, \binits{N.}},
\bauthor{\bsnm{Truhn}, \binits{D.}},
\bauthor{\bsnm{Veldhuizen}, \binits{G.P.}},
\bauthor{\bsnm{Han}, \binits{T.}},
\bauthor{\bsnm{Treeck}, \binits{M.}},
\bauthor{\bsnm{Buelow}, \binits{R.D.}},
\bauthor{\bsnm{Langer}, \binits{R.}},
\bauthor{\bsnm{Dislich}, \binits{B.}},
\bauthor{\bsnm{Boor}, \binits{P.}},
\bauthor{\bsnm{Schulz}, \binits{V.}}, \betal:
\batitle{Adversarial attacks and adversarial robustness in computational pathology}.
\bjtitle{Nature communications}
\bvolume{13}(\bissue{1}),
\bfpage{5711}
(\byear{2022})
\end{barticle}
\endbibitem

\bibitem[\protect\citeauthoryear{Kanwal and Engan}{2024}]{kanwal2024extract}
\begin{botherref}
\oauthor{\bsnm{Kanwal}, \binits{N.}},
\oauthor{\bsnm{Engan}, \binits{K.}}:
Extract, detect, eliminate: Enhancing reliability and performance of computational pathology through artifact processing pipelines.
Science Talks
(2024)
\end{botherref}
\endbibitem

\bibitem[\protect\citeauthoryear{Kothari et~al.}{2013}]{kothari2013eliminating}
\begin{barticle}
\bauthor{\bsnm{Kothari}, \binits{S.}},
\bauthor{\bsnm{Phan}, \binits{J.H.}},
\bauthor{\bsnm{Wang}, \binits{M.D.}}:
\batitle{Eliminating tissue-fold artifacts in histopathological whole-slide images for improved image-based prediction of cancer grade}.
\bjtitle{Journal of pathology informatics}
\bvolume{4}(\bissue{1}),
\bfpage{22}
(\byear{2013})
\end{barticle}
\endbibitem

\bibitem[\protect\citeauthoryear{Kanwal et~al.}{2024}]{kanwal2024sure}
\begin{barticle}
\bauthor{\bsnm{Kanwal}, \binits{N.}},
\bauthor{\bsnm{L{\'o}pez-P{\'e}rez}, \binits{M.}},
\bauthor{\bsnm{Kiraz}, \binits{U.}},
\bauthor{\bsnm{Zuiverloon}, \binits{T.C.}},
\bauthor{\bsnm{Molina}, \binits{R.}},
\bauthor{\bsnm{Engan}, \binits{K.}}:
\batitle{Are you sure it’s an artifact? artifact detection and uncertainty quantification in histological images}.
\bjtitle{Computerized Medical Imaging and Graphics}
\bvolume{112},
\bfpage{102321}
(\byear{2024})
\end{barticle}
\endbibitem

\bibitem[\protect\citeauthoryear{Salvi et~al.}{2021}]{salvi2021impact}
\begin{barticle}
\bauthor{\bsnm{Salvi}, \binits{M.}},
\bauthor{\bsnm{Acharya}, \binits{U.R.}},
\bauthor{\bsnm{Molinari}, \binits{F.}},
\bauthor{\bsnm{Meiburger}, \binits{K.M.}}:
\batitle{The impact of pre-and post-image processing techniques on deep learning frameworks: A comprehensive review for digital pathology image analysis}.
\bjtitle{Computers in Biology and Medicine}
\bvolume{128},
\bfpage{104129}
(\byear{2021})
\end{barticle}
\endbibitem

\bibitem[\protect\citeauthoryear{P{\'e}rez-Bueno et~al.}{2020}]{Perez-Bueno2020a}
\begin{bchapter}
\bauthor{\bsnm{P{\'e}rez-Bueno}, \binits{F.}},
\bauthor{\bsnm{Vega}, \binits{M.}},
\bauthor{\bsnm{Naranjo}, \binits{V.}},
\bauthor{\bsnm{Molina}, \binits{R.}},
\bauthor{\bsnm{Katsaggelos}, \binits{A.K.}}:
\bctitle{Super gaussian priors for blind color deconvolution of histological images}.
In: \bbtitle{2020 IEEE International Conference on Image Processing (ICIP)},
pp. \bfpage{3010}--\blpage{3014}
(\byear{2020}).
\bcomment{IEEE}
\end{bchapter}
\endbibitem

\bibitem[\protect\citeauthoryear{Ameisen et~al.}{2014}]{Ameisen2014}
\begin{barticle}
\bauthor{\bsnm{Ameisen}, \binits{D.}},
\bauthor{\bsnm{Deroulers}, \binits{C.}},
\bauthor{\bsnm{Perrier}, \binits{V.}},
\bauthor{\bsnm{Bouhidel}, \binits{F.}},
\bauthor{\bsnm{Battistella}, \binits{M.}},
\bauthor{\bsnm{Legr{\`{e}}s}, \binits{L.}},
\bauthor{\bsnm{Janin}, \binits{A.}},
\bauthor{\bsnm{Bertheau}, \binits{P.}},
\bauthor{\bsnm{Yun{\`{e}}s}, \binits{J.B.}}:
\batitle{{Towards better digital pathology workflows: Programming libraries for high-speed sharpness assessment of Whole Slide Images}}.
\bjtitle{Diagnostic Pathology}
\bvolume{9}(\bissue{1}),
\bfpage{1}--\blpage{7}
(\byear{2014})
\end{barticle}
\endbibitem

\bibitem[\protect\citeauthoryear{Shrestha et~al.}{2016}]{shrestha2016}
\begin{botherref}
\oauthor{\bsnm{Shrestha}, \binits{P.}},
\oauthor{\bsnm{Kneepkens}, \binits{R.}},
\oauthor{\bsnm{Vrijnsen}, \binits{J.}},
\oauthor{\bsnm{Vossen}, \binits{D.}},
\oauthor{\bsnm{Abels}, \binits{E.}},
\oauthor{\bsnm{Hulsken}, \binits{B.}}:
A quantitative approach to evaluate image quality of whole slide imaging scanners.
Journal of pathology informatics
\textbf{7}
(2016)
\end{botherref}
\endbibitem

\bibitem[\protect\citeauthoryear{Bahlmann et~al.}{2012}]{bahlmann2012automated}
\begin{bchapter}
\bauthor{\bsnm{Bahlmann}, \binits{C.}},
\bauthor{\bsnm{Patel}, \binits{A.}},
\bauthor{\bsnm{Johnson}, \binits{J.}},
\bauthor{\bsnm{Ni}, \binits{J.}},
\bauthor{\bsnm{Chekkoury}, \binits{A.}},
\bauthor{\bsnm{Khurd}, \binits{P.}},
\bauthor{\bsnm{Kamen}, \binits{A.}},
\bauthor{\bsnm{Grady}, \binits{L.}},
\bauthor{\bsnm{Krupinski}, \binits{E.}},
\bauthor{\bsnm{Graham}, \binits{A.}}, \betal:
\bctitle{Automated detection of diagnostically relevant regions in h\&e stained digital pathology slides}.
In: \bbtitle{Medical Imaging 2012: Computer-Aided Diagnosis},
vol. \bseriesno{8315},
p. \bfpage{831504}
(\byear{2012}).
\bcomment{International Society for Optics and Photonics}
\end{bchapter}
\endbibitem

\bibitem[\protect\citeauthoryear{Avanaki et~al.}{2016}]{Avanaki2016}
\begin{bchapter}
\bauthor{\bsnm{Avanaki}, \binits{A.R.N.}},
\bauthor{\bsnm{Espig}, \binits{K.S.}},
\bauthor{\bsnm{Xthona}, \binits{A.}},
\bauthor{\bsnm{Lanciault}, \binits{C.}},
\bauthor{\bsnm{Kimpe}, \binits{T.R.L.}}:
\bctitle{Automatic image quality assessment for digital pathology}.
In: \beditor{\bsnm{Tingberg}, \binits{A.}},
\beditor{\bsnm{L{\aa}ng}, \binits{K.}},
\beditor{\bsnm{Timberg}, \binits{P.}} (eds.)
\bbtitle{Breast Imaging},
pp. \bfpage{431}--\blpage{438}.
\bpublisher{Springer},
\blocation{Cham}
(\byear{2016})
\end{bchapter}
\endbibitem

\bibitem[\protect\citeauthoryear{Janowczyk et~al.}{2019}]{histoqc}
\begin{barticle}
\bauthor{\bsnm{Janowczyk}, \binits{A.}},
\bauthor{\bsnm{Zuo}, \binits{R.}},
\bauthor{\bsnm{Gilmore}, \binits{H.}},
\bauthor{\bsnm{Feldman}, \binits{M.}},
\bauthor{\bsnm{Madabhushi}, \binits{A.}}:
\batitle{Histoqc: an open-source quality control tool for digital pathology slides}.
\bjtitle{JCO clinical cancer informatics}
\bvolume{3},
\bfpage{1}--\blpage{7}
(\byear{2019})
\end{barticle}
\endbibitem

\bibitem[\protect\citeauthoryear{Gao et~al.}{2010}]{gao2010automated}
\begin{bchapter}
\bauthor{\bsnm{Gao}, \binits{D.}},
\bauthor{\bsnm{Padfield}, \binits{D.}},
\bauthor{\bsnm{Rittscher}, \binits{J.}},
\bauthor{\bsnm{McKay}, \binits{R.}}:
\bctitle{Automated training data generation for microscopy focus classification}.
In: \bbtitle{International Conference on Medical Image Computing and Computer-Assisted Intervention},
pp. \bfpage{446}--\blpage{453}
(\byear{2010}).
\bcomment{Springer}
\end{bchapter}
\endbibitem

\bibitem[\protect\citeauthoryear{Hashimoto et~al.}{2012}]{hashimoto2012}
\begin{botherref}
\oauthor{\bsnm{Hashimoto}, \binits{N.}},
\oauthor{\bsnm{Bautista}, \binits{P.A.}},
\oauthor{\bsnm{Yamaguchi}, \binits{M.}},
\oauthor{\bsnm{Ohyama}, \binits{N.}},
\oauthor{\bsnm{Yagi}, \binits{Y.}}:
Referenceless image quality evaluation for whole slide imaging.
Journal of pathology informatics
\textbf{3}
(2012)
\end{botherref}
\endbibitem

\bibitem[\protect\citeauthoryear{Palokangas et~al.}{2007}]{palokangas2007segmentation}
\begin{bchapter}
\bauthor{\bsnm{Palokangas}, \binits{S.}},
\bauthor{\bsnm{Selinummi}, \binits{J.}},
\bauthor{\bsnm{Yli-Harja}, \binits{O.}}:
\bctitle{Segmentation of folds in tissue section images}.
In: \bbtitle{2007 29th Annual International Conference of the IEEE Engineering in Medicine and Biology Society},
pp. \bfpage{5641}--\blpage{5644}
(\byear{2007}).
\bcomment{IEEE}
\end{bchapter}
\endbibitem

\bibitem[\protect\citeauthoryear{Bautista and Yagi}{2009}]{Bautista2009}
\begin{botherref}
\oauthor{\bsnm{Bautista}, \binits{P.A.}},
\oauthor{\bsnm{Yagi}, \binits{Y.}}:
{Detection of tissue folds in whole slide images}.
Proceedings of the 31st Annual International Conference of the IEEE Engineering in Medicine and Biology Society: Engineering the Future of Biomedicine, EMBC 2009,
3669--3672
(2009)
\end{botherref}
\endbibitem

\bibitem[\protect\citeauthoryear{Swiderska-Chadaj et~al.}{2016}]{swiderska2016automatic}
\begin{bchapter}
\bauthor{\bsnm{Swiderska-Chadaj}, \binits{Z.}},
\bauthor{\bsnm{Markiewicz}, \binits{T.}},
\bauthor{\bsnm{Cierniak}, \binits{S.}},
\bauthor{\bsnm{Koktysz}, \binits{R.}}:
\bctitle{Automatic quantification of vessels in hemorrhoids whole slide images}.
In: \bbtitle{2016 17th International Conference Computational Problems of Electrical Engineering (CPEE)},
pp. \bfpage{1}--\blpage{4}
(\byear{2016}).
\bcomment{IEEE}
\end{bchapter}
\endbibitem

\bibitem[\protect\citeauthoryear{Mercan et~al.}{2014}]{mercan2014localization}
\begin{bchapter}
\bauthor{\bsnm{Mercan}, \binits{E.}},
\bauthor{\bsnm{Aksoy}, \binits{S.}},
\bauthor{\bsnm{Shapiro}, \binits{L.G.}},
\bauthor{\bsnm{Weaver}, \binits{D.L.}},
\bauthor{\bsnm{Brunye}, \binits{T.}},
\bauthor{\bsnm{Elmore}, \binits{J.G.}}:
\bctitle{Localization of diagnostically relevant regions of interest in whole slide images}.
In: \bbtitle{2014 22nd International Conference on Pattern Recognition},
pp. \bfpage{1179}--\blpage{1184}
(\byear{2014}).
\bcomment{IEEE}
\end{bchapter}
\endbibitem

\bibitem[\protect\citeauthoryear{Albuquerque et~al.}{2021}]{albuquerque2021deep}
\begin{bchapter}
\bauthor{\bsnm{Albuquerque}, \binits{T.}},
\bauthor{\bsnm{Moreira}, \binits{A.}},
\bauthor{\bsnm{Cardoso}, \binits{J.S.}}:
\bctitle{Deep ordinal focus assessment for whole slide images}.
In: \bbtitle{Proceedings of the IEEE/CVF International Conference on Computer Vision},
pp. \bfpage{657}--\blpage{663}
(\byear{2021})
\end{bchapter}
\endbibitem

\bibitem[\protect\citeauthoryear{Kohlberger et~al.}{2019}]{kohlberger2019whole}
\begin{botherref}
\oauthor{\bsnm{Kohlberger}, \binits{T.}},
\oauthor{\bsnm{Liu}, \binits{Y.}},
\oauthor{\bsnm{Moran}, \binits{M.}},
\oauthor{\bsnm{Chen}, \binits{P.-H.C.}},
\oauthor{\bsnm{Brown}, \binits{T.}},
\oauthor{\bsnm{Hipp}, \binits{J.D.}},
\oauthor{\bsnm{Mermel}, \binits{C.H.}},
\oauthor{\bsnm{Stumpe}, \binits{M.C.}}:
Whole-slide image focus quality: Automatic assessment and impact on ai cancer detection.
Journal of pathology informatics
\textbf{10}
(2019)
\end{botherref}
\endbibitem

\bibitem[\protect\citeauthoryear{Wetteland et~al.}{2019}]{Wetteland2019}
\begin{botherref}
\oauthor{\bsnm{Wetteland}, \binits{R.}},
\oauthor{\bsnm{Engan}, \binits{K.}},
\oauthor{\bsnm{Eftest{\o}l}, \binits{T.}},
\oauthor{\bsnm{Kvikstad}, \binits{V.}},
\oauthor{\bsnm{Janssen}, \binits{E.A.M.}}:
{Multiclass tissue classification of whole-slide histological images using convolutional neural networks}.
ICPRAM 2019 - Proceedings of the 8th International Conference on Pattern Recognition Applications and Methods,
320--327
(2019)
\end{botherref}
\endbibitem

\bibitem[\protect\citeauthoryear{Wetteland et~al.}{2020}]{wetteland2020multiscale}
\begin{barticle}
\bauthor{\bsnm{Wetteland}, \binits{R.}},
\bauthor{\bsnm{Engan}, \binits{K.}},
\bauthor{\bsnm{Eftest{\o}l}, \binits{T.}},
\bauthor{\bsnm{Kvikstad}, \binits{V.}},
\bauthor{\bsnm{Janssen}, \binits{E.A.}}:
\batitle{A multiscale approach for whole-slide image segmentation of five tissue classes in urothelial carcinoma slides}.
\bjtitle{Technology in Cancer Research \& Treatment}
\bvolume{19},
\bfpage{1533033820946787}
(\byear{2020})
\end{barticle}
\endbibitem

\bibitem[\protect\citeauthoryear{Clymer et~al.}{2020}]{clymer2020decidual}
\begin{barticle}
\bauthor{\bsnm{Clymer}, \binits{D.}},
\bauthor{\bsnm{Kostadinov}, \binits{S.}},
\bauthor{\bsnm{Catov}, \binits{J.}},
\bauthor{\bsnm{Skvarca}, \binits{L.}},
\bauthor{\bsnm{Pantanowitz}, \binits{L.}},
\bauthor{\bsnm{Cagan}, \binits{J.}},
\bauthor{\bsnm{LeDuc}, \binits{P.}}:
\batitle{Decidual vasculopathy identification in whole slide images using multiresolution hierarchical convolutional neural networks}.
\bjtitle{The American Journal of Pathology}
\bvolume{190}(\bissue{10}),
\bfpage{2111}--\blpage{2122}
(\byear{2020})
\end{barticle}
\endbibitem

\bibitem[\protect\citeauthoryear{Babaie and Tizhoosh}{2019}]{babaie2019deep}
\begin{bchapter}
\bauthor{\bsnm{Babaie}, \binits{M.}},
\bauthor{\bsnm{Tizhoosh}, \binits{H.R.}}:
\bctitle{Deep features for tissue-fold detection in histopathology images}.
In: \bbtitle{European Congress on Digital Pathology},
pp. \bfpage{125}--\blpage{132}
(\byear{2019}).
\bcomment{Springer}
\end{bchapter}
\endbibitem

\bibitem[\protect\citeauthoryear{Kanwal et~al.}{2022}]{kanwal2022quantifying}
\begin{bchapter}
\bauthor{\bsnm{Kanwal}, \binits{N.}},
\bauthor{\bsnm{Fuster}, \binits{S.}},
\bauthor{\bsnm{Khoraminia}, \binits{F.}},
\bauthor{\bsnm{Zuiverloon}, \binits{T.C.}},
\bauthor{\bsnm{Rong}, \binits{C.}},
\bauthor{\bsnm{Engan}, \binits{K.}}:
\bctitle{Quantifying the effect of color processing on blood and damaged tissue detection in whole slide images}.
In: \bbtitle{2022 IEEE 14th Image, Video, and Multidimensional Signal Processing Workshop (IVMSP)},
pp. \bfpage{1}--\blpage{5}
(\byear{2022}).
\bcomment{IEEE}
\end{bchapter}
\endbibitem

\bibitem[\protect\citeauthoryear{Guo et~al.}{2017}]{guo2017calibration}
\begin{bchapter}
\bauthor{\bsnm{Guo}, \binits{C.}},
\bauthor{\bsnm{Pleiss}, \binits{G.}},
\bauthor{\bsnm{Sun}, \binits{Y.}},
\bauthor{\bsnm{Weinberger}, \binits{K.Q.}}:
\bctitle{On calibration of modern neural networks}.
In: \bbtitle{International Conference on Machine Learning},
pp. \bfpage{1321}--\blpage{1330}
(\byear{2017}).
\bcomment{PMLR}
\end{bchapter}
\endbibitem

\bibitem[\protect\citeauthoryear{Linmans et~al.}{2023}]{linmans2023predictive}
\begin{barticle}
\bauthor{\bsnm{Linmans}, \binits{J.}},
\bauthor{\bsnm{Elfwing}, \binits{S.}},
\bauthor{\bsnm{Laak}, \binits{J.}},
\bauthor{\bsnm{Litjens}, \binits{G.}}:
\batitle{Predictive uncertainty estimation for out-of-distribution detection in digital pathology}.
\bjtitle{Medical Image Analysis}
\bvolume{83},
\bfpage{102655}
(\byear{2023})
\end{barticle}
\endbibitem

\bibitem[\protect\citeauthoryear{Zhang et~al.}{2021}]{zhang2021understanding}
\begin{barticle}
\bauthor{\bsnm{Zhang}, \binits{C.}},
\bauthor{\bsnm{Bengio}, \binits{S.}},
\bauthor{\bsnm{Hardt}, \binits{M.}},
\bauthor{\bsnm{Recht}, \binits{B.}},
\bauthor{\bsnm{Vinyals}, \binits{O.}}:
\batitle{Understanding deep learning (still) requires rethinking generalization}.
\bjtitle{Communications of the ACM}
\bvolume{64}(\bissue{3}),
\bfpage{107}--\blpage{115}
(\byear{2021})
\end{barticle}
\endbibitem

\bibitem[\protect\citeauthoryear{Dosovitskiy et~al.}{2020}]{dosovitskiy2020image}
\begin{botherref}
\oauthor{\bsnm{Dosovitskiy}, \binits{A.}},
\oauthor{\bsnm{Beyer}, \binits{L.}},
\oauthor{\bsnm{Kolesnikov}, \binits{A.}},
\oauthor{\bsnm{Weissenborn}, \binits{D.}},
\oauthor{\bsnm{Zhai}, \binits{X.}},
\oauthor{\bsnm{Unterthiner}, \binits{T.}},
\oauthor{\bsnm{Dehghani}, \binits{M.}},
\oauthor{\bsnm{Minderer}, \binits{M.}},
\oauthor{\bsnm{Heigold}, \binits{G.}},
\oauthor{\bsnm{Gelly}, \binits{S.}}, et al.:
An image is worth 16x16 words: Transformers for image recognition at scale.
arXiv preprint arXiv:2010.11929
(2020)
\end{botherref}
\endbibitem

\bibitem[\protect\citeauthoryear{Deng et~al.}{2009}]{imagenet}
\begin{bchapter}
\bauthor{\bsnm{Deng}, \binits{J.}},
\bauthor{\bsnm{Dong}, \binits{W.}},
\bauthor{\bsnm{Socher}, \binits{R.}},
\bauthor{\bsnm{Li}, \binits{L.-J.}},
\bauthor{\bsnm{Li}, \binits{K.}},
\bauthor{\bsnm{Fei-Fei}, \binits{L.}}:
\bctitle{Imagenet: A large-scale hierarchical image database}.
In: \bbtitle{2009 IEEE ICCV},
pp. \bfpage{248}--\blpage{255}
(\byear{2009}).
\bcomment{Ieee}
\end{bchapter}
\endbibitem

\bibitem[\protect\citeauthoryear{Wetzer}{2023}]{wetzer2023representation}
\begin{botherref}
\oauthor{\bsnm{Wetzer}, \binits{E.}}:
Representation learning and information fusion: Applications in biomedical image processing.
PhD thesis,
Acta Universitatis Upsaliensis
(2023)
\end{botherref}
\endbibitem

\bibitem[\protect\citeauthoryear{Shakhawat et~al.}{2020}]{Shakhawat2020}
\begin{barticle}
\bauthor{\bsnm{Shakhawat}, \binits{H.M.}},
\bauthor{\bsnm{Nakamura}, \binits{T.}},
\bauthor{\bsnm{Kimura}, \binits{F.}},
\bauthor{\bsnm{Yagi}, \binits{Y.}},
\bauthor{\bsnm{Yamaguchi}, \binits{M.}}:
\batitle{{Automatic quality evaluation of whole slide images for the practical use of whole slide imaging scanner}}.
\bjtitle{ITE Transactions on Media Technology and Applications}
\bvolume{8}(\bissue{4}),
\bfpage{252}--\blpage{268}
(\byear{2020})
\end{barticle}
\endbibitem

\bibitem[\protect\citeauthoryear{Senaras et~al.}{2018}]{senaras2018deepfocus}
\begin{barticle}
\bauthor{\bsnm{Senaras}, \binits{C.}},
\bauthor{\bsnm{Niazi}, \binits{M.K.K.}},
\bauthor{\bsnm{Lozanski}, \binits{G.}},
\bauthor{\bsnm{Gurcan}, \binits{M.N.}}:
\batitle{Deepfocus: detection of out-of-focus regions in whole slide digital images using deep learning}.
\bjtitle{PloS one}
\bvolume{13}(\bissue{10}),
\bfpage{0205387}
(\byear{2018})
\end{barticle}
\endbibitem

\bibitem[\protect\citeauthoryear{Raipuria and Singhal}{2022}]{raipuria2022stress}
\begin{bchapter}
\bauthor{\bsnm{Raipuria}, \binits{G.}},
\bauthor{\bsnm{Singhal}, \binits{N.}}:
\bctitle{Stress testing vision transformers using common histopathological artifacts}.
In: \bbtitle{Medical Imaging with Deep Learning}
(\byear{2022})
\end{bchapter}
\endbibitem

\bibitem[\protect\citeauthoryear{Swiderska-Chadaj et~al.}{2018}]{swiderska2018deep}
\begin{botherref}
\oauthor{\bsnm{Swiderska-Chadaj}, \binits{Z.}},
\oauthor{\bsnm{Markiewicz}, \binits{T.}},
\oauthor{\bsnm{Gallego}, \binits{J.}},
\oauthor{\bsnm{Bueno}, \binits{G.}},
\oauthor{\bsnm{Grala}, \binits{B.}},
\oauthor{\bsnm{Lorent}, \binits{M.}}:
Deep learning for damaged tissue detection and segmentation in ki-67 brain tumor specimens based on the u-net model.
Bulletin of the Polish Academy of Sciences: Technical Sciences,
849--856
(2018)
\end{botherref}
\endbibitem

\end{thebibliography}
\end{document}